\documentclass[a4paper,11pt]{article}
\pdfoutput=1 % if your are submitting a pdflatex (i.e. if you have
\usepackage{jcappub} % for details on the use of the package, please
\usepackage[T1]{fontenc} % if needed

\title{ %Evolving black holes in a nonsingular bouncing universe
       Evolution of black holes through a nonsingular cosmological bounce 
}

\author[a,b]{Maxence Corman}
\author[a]{William E. East}
\author[c]{Justin L. Ripley}

\affiliation[a]{
   Perimeter Institute for Theoretical Physics,
   Waterloo, Ontario N2L 2Y5, Canada.
}
\affiliation[b]{
   Department of Physics and Astronomy, University of Waterloo, Waterloo, Ontario N2L 3G1,
Canada
}
\affiliation[c]{
   DAMTP, Centre for Mathematical Sciences,
   University of Cambridge,
   Wilberforce Road,
   Cambridge CB3 0WA,
   United Kingdom
}
\emailAdd{mcorman@perimeterinstitute.ca}
\emailAdd{weast@perimeterinstitute.ca}
\emailAdd{jr860@cam.ac.uk}

\abstract{
   We study the  
   classical dynamics of black holes during a nonsingular cosmological bounce.
   Taking a simple model of a nonsingular bouncing cosmology driven by the combination of a ghost and ordinary
   scalar field, we use nonlinear evolutions of the Einstein equations to follow rotating and non-rotating
   black holes of different sizes through the bounce.
   The violation of the null energy condition allows for a shrinking black hole event horizon and
   we find that for sufficiently large black holes (relative to the minimum Hubble radius)
   the black hole apparent horizon can disappear during the contraction phase.
   Despite this, we show that most of the local cosmological 
   evolution remains largely unaffected by the presence
   of the black hole.
   We find that, independently of the black hole's initial mass, 
   the black hole's event horizon persists throughout the bounce,
   and the late time dynamics consists of an expanding universe 
   with a black hole of mass comparable to its initial value. 
}

\begin{document}
\maketitle
\flushbottom

%=============================================================================
\section{Introduction\label{sec:intro}}
A proposed alternative to cosmic inflation is the idea
that the universe underwent a bounce: a transition from a stage of contraction to expansion
\cite{Gasperini:2002bn,Brandenberger:2012zb,Novello:2008ra,Battefeld:2014uga,
Brandenberger:2016vhg}.
In a singular bounce, the universe passes through a classical 
singularity where the cosmological scale factor
becomes small, curvature invariants blow up, and quantum gravity effects
presumably become highly relevant to determining the future dynamics of the
universe \cite{Tolley:2003nx,McFadden:2005mq,
Bars:2011aa,Gielen:2015uaa}.
An alternative, which we focus on here, is a nonsingular bounce.  For such
cosmologies, so long as the spacetime curvature does not become Planckian,
there is the possibility that quantum gravity effects could be subdominant to
classical effects, in which case one may be able to describe the dynamics of
the bounce using classical physics.  Nonsingular bouncing cosmologies require
violating the null convergence condition (NCC), which states that for all null
vectors $k^{\mu}$, $R_{\mu\nu}k^{\mu}k^{\nu}\geq0$
\cite{Molina-Paris:1998xmn,Khoury:2001bz,
Battefeld:2014uga,Rubakov:2014jja,Brandenberger:2016vhg}. In Einstein gravity,
the NCC is equivalent to the null energy condition which is satisfied by most
standard classical field theories \cite{Rubakov:2014jja}. Nonsingular bouncing
cosmologies hence require non-standard matter terms, or modifications to
Einstein gravity, for example, Horndeski theories including ghost condensation
\cite{Arkani-Hamed:2003pdi,Buchbinder:2007ad} or (cubic) Galileon/Horndeski
models
\cite{Dubovsky:2005xd,Easson:2011zy,Cai:2012va,
Elder:2013gya,Ijjas:2016vtq,Ijjas:2016tpn}.
While perturbative studies of these theories suggest they may be free of ghost
or gradient instabilities \cite{Easson:2011zy,Ijjas:2016vtq}, less
is known about which models will remain (strongly) hyperbolic through a bounce,
when the solution is presumably not in the weakly coupled regime
\cite{Papallo:2017qvl,Kovacs:2020pns,Kovacs:2020ywu}\footnote{We
note that the model proposed in \cite{Ijjas:2016vtq} is known to
break down shortly after the bounce has ended \cite{Dobre:2017pnt}.}

An important open question is what happens in bouncing cosmologies in the
inhomogeneous and non-perturbative regime. While there are several analytical
and numerical studies of the dynamics of bouncing cosmologies during their
contraction phase
\cite{Garfinkle:2008ei,Ijjas:2020dws,Cook:2020oaj,Ijjas:2021wml}, there are
relatively few studies of the dynamics of the bounce
\cite{Peter:2002cn,Allen:2004vz,Cai:2007zv,Xue:2013bva,Ijjas:2016vtq}, and none
that consider the dynamics of black holes beyond the restriction to spherical
symmetry~\cite{faraoni2015cosmological}. 
Previous studies of black hole--cosmological bounces
have either constructed initial data for black hole bouncing solutions 
\cite{Clifton:2017hvg}, worked in a perturbative limit 
\cite{Carr:2011hv,Chen:2016kjx,Ijjas:2021zwv,Gorkavyi:2021hmb},
or made use of analytic solutions 
(e.g. generalizations of the McVittie solutions 
\cite{Perez:2021etn,Perez:2021yht,Perez:2022hvy}), 
that are limited by the fact that the metric evolution is prescribed 
ad-hoc, and from that the implied matter type and evolution is derived.
The question of what happens to a
black hole in a nonsingular cosmological bounce is particular salient for
several reasons. On the one hand, the bounce necessarily requires a violation
of the assumptions made in black hole singularity theorems and results on black
hole horizons (namely the NCC)~\cite{hawking1973large}, 
so there is a question of whether
the black hole will survive the bounce, or if the bounce mechanism will
also reverse gravitational collapse, and if this will possibly lead to a naked
singularity. On the other hand, one might also worry what the backreaction of
the black hole's gravity will be on the bounce in the neighbourhood of the
black hole.  
An extreme scenario would be if the bounce failed to happen in the 
vicinity of the black hole, possibly leading to a patch of
contraction that grows into the expanding spacetime, as happens, e.g., in
scenarios where the Higgs boson
is destabilized during inflation, and goes to its
true vacuum at negative energy 
densities~\cite{Espinosa:2015qea,East:2016anr}. 

Here, we address these questions by studying the nonlinear
dynamics and evolution of black holes in a particular nonsingular bouncing cosmology
(details of which are described below).
Black holes can be expected to form during the contraction of
matter and radiation dominated universes 
\cite{Banks:2002fe,Chen:2016kjx,Quintin:2016qro}, and will generally
be present from previous eras in cyclic cosmologies~\cite{Steinhardt:2001st,Ijjas:2019pyf}.
However, it is common to invoke a smoothing phase during contraction
(e.g. ekpyrosis~\cite{Khoury:2001wf,Steinhardt:2001st,Erickson:2003zm}),
and argue that Hubble patches containing a black hole will be rare. 
Regardless, we view our work as serving two main purposes:
(1) to study the dynamics and robustness of a nonsingular bouncing model when
a very large perturbation, namely a black hole, is introduced, and
(2) to explore the dynamics of the black hole and cosmological horizons 
during the bounce.

To avoid the difficulties related to finding a motivated theory that can give rise to bouncing
solutions while also having well-posed evolution equations in the inhomogeneous regime, 
and thus being suitable for describing
black hole dynamics, we will work with a bouncing cosmology model that incorporates a 
minimally coupled scalar field 
with a ghost field (i.e. a field which contributes to a negative cosmological
energy density), to drive the bounce. 
While ghost fields are known to give problematic quantum mechanical theories
(for a discussion of this in the context of cosmology see \cite{Cline:2003gs,Kallosh:2007ad};
see also \cite{Adams:2006sv}), 
we take the point of view of \cite{Peter:2002cn,Allen:2004vz,Xue:2013bva} 
and treat the ghost field as an \emph{effective} model for NCC violation.
Quantum stability and unitarity is a distinct issue requiring a separate
analysis (see, e.g.,~\cite{deRham:2017aoj}).
Unlike earlier work with this model, 
we do not restrict ourselves to cosmological
spacetimes that have planar symmetry~\cite{Xue:2013bva}, 
or to small linear perturbations about a background bouncing spacetime
\cite{Peter:2002cn,Allen:2004vz}.
Instead, we consider contracting cosmological initial data that contains
a black hole, and work in an axisymmetric spacetime.
This allows us to examine the effect that a
large inhomogeneity has on 
the dynamics of the spacetime near and during the bounce.

Following the growing number of studies making use of techniques from 
numerical relativity to study cosmological phenomena involving
black hole dynamics~\cite{Bentivegna:2012ei,Yoo:2013yea,
East:2015ggf,East:2016anr,Clough:2016ymm,Aurrekoetxea:2019fhr,
Giblin:2019pql,Joana:2020rxm,Corman:2021osa}, 
we use numerical solutions to follow the evolution of different
size black holes, both non-spinning and spinning, through a bounce, considering
those both bigger and smaller than the minimum Hubble radius. 
Our main results are that the black holes persist to the expanding phase, and that
the nonsingular bouncing model under study is fairly robust
under large perturbations, in the sense that the local spacetime expansion
around the black hole successfully bounces for all of the cases we explored.
For large enough black holes, we find the black hole apparent horizon
collides with the cosmological horizon, and temporarily disappears during the contraction
phase. Nevertheless, the black hole apparent horizon eventually reappears (with
finite radius event horizon throughout) and this does not disrupt the bounce at late times. 

In principle a nonsingular, 
\emph{classical} bounce could occur at any characteristic length scale 
that is larger than the scale at which quantum gravity effects
become important (presumably the Planck scale:
$l_P\sim 10^{-33}$ cm in geometric units).
Given this, the length scale of a classical nonsingular bounce can still
be extremely small compared to the typical length scale of say,
an astrophysical black hole (e.g. in \cite{Ijjas:2019pyf} the bounce
happens at a typical length scale of $\sim 10^{-25}\mathrm{cm}\sim10^8 l_P$).
One may expect then that if any Hubble patch were to contain a black hole, 
that the black hole would be much larger than
the minimum size of the Hubble patch.
For example, even a black hole with a mass of
$m_{BH}\sim 10^{15}$ g\footnote{Primordial black hole with masses
smaller than $m_{BH}\sim 10^{15}$ g would have evaporated by now due to Hawking evaporation; 
this is then a reasonable lower bound
for the mass of black holes that were present 
in the early universe \cite{CHAPLINE1975,Carr:2020gox}.}
at the bounce would still have a size of $\sim 10^{20}l_P$;
this is orders of magnitude larger 
than the example bounce scale mentioned above.
For this reason, we will be more interested in considering black holes
whose size is comparable or larger than the bounce scale
(which we take to be $1/|H_{\rm min}|$, where $H_{\rm min}<0$ is the maximum
contraction rate).

The remainder of this paper is as follows. 
We discuss the nonsingular bouncing model we use in section~\ref{sec:ghost_field}. 
Our numerical methods and diagnostics for evolving the 
nonsingular bounce are outlined in 
section~\ref{sec:diagnostics}. 
Our numerical results are described in 
section~\ref{sec:numerical_results},
and we conclude in section~\ref{sec:discussion}.
In appendix ~\ref{sec:numerical_methodology},
we discuss our numerical methodology in more detail,
in appendix ~\ref{sec:DH}, we define various quasi-local notions of black hole
and cosmological horizons, and
in appendix ~\ref{sec:mcvittie},
we provide an overview of the McVittie spacetime,
an analytic solution to the Einstein equations of a black hole embedded
in a cosmology, of which our numerical simulations can be seen as
a generalization.
%-----------------------------------------------------------------------------
\subsection{Conventions and notation}
\label{sec:conventions_notation}
   We work in four spacetime dimensions, with metric signature
$(-+++)$; we use lower-case Greek letters ($\mu,\nu,...$)
to denote spacetime indices and Latin letters ($i,j,k,...$, although $t$
is reserved for the time coordinate index)
to denote spatial indices. 
The Riemann tensor is
$R^{\alpha}{}_{\beta\gamma\delta}
=
\partial_{\gamma}\Gamma^{\alpha}_{\beta\delta}-\cdots
$.
We use units with $G=c=\hbar=1$.
%=============================================================================
\section{Ghost field model}\label{sec:ghost_field}

We consider a theory that has two scalar fields $\phi$ and $\chi$ coupled
to gravity:
\begin{align}
   \label{eq:act_ghost}
   S
   =
   \int d^4x\sqrt{-g}\left(\frac{1}{16\pi}
      R
   -  \nabla_{\alpha}\phi\nabla^{\alpha}\phi 
   -  2V\left(\phi\right)
   +  \nabla_{\alpha}\chi\nabla^{\alpha}\chi 
   \right)
   .
\end{align}
This model has a canonically normalized scalar field $\phi$
with a potential $V(\phi)=V_0 e^{-c \phi}$,
and a massless ghost field $\chi$.

The covariant equations of motion for \eqref{eq:act_ghost} are 
\begin{subequations}
\label{eq:equations_of_motion}
\begin{align}
    \nabla^{\alpha} \nabla_{\alpha} \phi -\frac{dV}{d\phi}
   = &
   0
   ,\\
   \nabla^{\alpha} \nabla_{\alpha} \chi 
   = &
   0
   ,\\
    \frac{1}{8\pi}\left(
         R_{\alpha\beta}
      -  \frac{1}{2}g_{\alpha\beta}R
      \right) 
   +
   2 \nabla_{\alpha} \chi \nabla_{\beta} \chi
   -
   2 \nabla_{\alpha} \phi \nabla_{\beta} \phi
   + & \nonumber \\
   g_{\alpha\beta}  \left(
      2 V(\phi)
      +
      \nabla_c\phi \nabla^c \phi
      - 
      \nabla_c \chi \nabla^c \chi
   \right) 
   =&
   0
   .
\end{align}
\end{subequations}

Nonlinear, inhomogeneous cosmological solutions to the model \eqref{eq:act_ghost} 
were studied in \cite{Xue:2013bva}.
There, the authors considered a toroidal universe with a planar
perturbation in one of the spatial directions.
In this work, we consider an \emph{asymptotically} bouncing
FLRW universe with an initial black hole;
see section~\ref{sec:diagnostics} and 
appendix \ref{sec:numerical_methodology} for more details
on our numerical methodology.

Strictly speaking, the ghost field should be stabilized by 
some mechanism at the quantum level. 
We choose to ignore this and treat \eqref{eq:act_ghost} 
as a purely classical theory. 
As the equations of motion \eqref{eq:equations_of_motion}
have a well-posed initial value problem\footnote{More specifically,
the equations of motion form a strongly hyperbolic system
when written in the generalized harmonic formulation we employ
in our code.
},
we expect the model should admit at least short-time classical
solutions from generic initial data.
%=============================================================================
\subsection{Homogeneous bouncing cosmology}{\label{sec:equations_of_motion}

Here we briefly review homogeneous, isotropic bouncing solutions 
for the system \eqref{eq:equations_of_motion}
(see also \cite{Xue:2013bva,Allen:2004vz}), and discuss
the values used for our asymptotic initial data.
We work with harmonic coordinates
($g^{\mu\nu}\Gamma^{\alpha}_{\mu\nu}=0$), 
so that the metric line element is
\begin{align}
   \label{eq:line_element_harmonic_cosmo}
   ds^2
   =
   - 
   a(t)^6dt^2
   +
   a(t)^2\delta_{ij}dx^idx^j
   .
\end{align}
The scalar field equations and Friedmann equations are then  
\begin{subequations}
\label{eq:harm_ghost}
\begin{align}
\label{eq:flrw_phi_evolution}
   {\phi}''
   =&
     - a^6 V_{, \phi}
   ,\\
\label{eq:flrw_chi_evolution}
   {\chi}''
   =&
   0
   ,\\
\label{eq:flrw_ddota_evolution}
   \mathcal{H}' 
   =&
   16 \pi a^6 V\left(\phi\right) 
   ,\\
\label{eq:flrw_hamiltonian_constraint}
   \mathcal{H}^2  
   =& \frac{8 \pi}{3}\left(
   {\phi '}^2
   -
   {\chi'}^2
   +
   2 a^6 V\left(\phi\right)\right)
   .
\end{align}
\end{subequations}
where the $'$ is the derivative with respect
to the harmonic time coordinate $t$ related to proper time by 
$d \tau \equiv a^3 dt$,
and $\mathcal{H}$ is the harmonic Hubble parameter 
\begin{align}
   \mathcal{H} 
   \equiv 
   \frac{a'}{a} 
   \equiv 
   a^3H
   ,
\end{align}
where $H$ is the Hubble parameter defined with respect to the 
proper time $H\equiv (da/d\tau)/a$.
We define effective energy densities $\rho$ and pressures $P$
for the two scalar fields:
\begin{align}\label{eq:flrw_energies}
   \rho_{\phi}
   \equiv
   {\dot{\phi}}^2
   +
   2 V
   ,\qquad
   \rho_{\chi}
   \equiv
   -
   {\dot{\chi}}^2
   ,\qquad
   P_{\phi}
   \equiv
   {\dot{\phi}}^2
   -
   2 V
   ,\qquad
   P_{\chi}
   =
   -
   {\dot{\chi}}^2
   ,
\end{align}
where $\dot{f}\equiv df/d\tau$. 
The total effective equation of state is
\begin{align}
   w
   =
   \frac{P_{\phi}+P_{\chi}}{\rho_{\phi}+\rho_{\chi}}
   =
   -
   1
   +
   \frac{16 \pi}{3 H^2}\left({\dot{\phi}}^2-{\dot{\chi}}^2\right)
   ,
\end{align}
so $w < -1$ if $|\dot{\chi}|>|\dot{\phi}|$.
A requirement for having a nonsingular bounce is that $w<-1$, which
coincides with violation of the NCC\footnote{If we write
the tensor equations of motion as 
$R_{\alpha\beta}-\frac{1}{2}g_{\alpha\beta}R=8\pi T_{\alpha\beta}$,
then the NCC coincides with the Null Energy Condition 
for the stress-energy tensor $T_{\alpha\beta}$
\cite{hawking1973large}.}.
For example, if we consider the null vector 
$k^{\mu}\partial_{\mu} 
\equiv 
\left((1/a^3)\partial_t + (1/a)\partial_x\right)/\sqrt{2}$, 
we then have
\begin{align}
   R_{\mu\nu}k^{\mu}k^{\nu}
   =
   -
   \dot{H}
   \;\;\;
   \left(
      =
      8\pi \left(\dot{\phi}^2 - \dot{\chi}^2\right)
   \right)
   .
\end{align}
When the NCC holds, we see that $\dot{H}<0$, so that we have
cosmic deceleration during expansion ($H>0$), 
or cosmic acceleration during contraction ($H<0$). 
When the NCC is violated, $\dot{H}>0$, and cosmic contraction can be
slowed down, and even reversed to make a bounce.

We also define effective equations of state for the fields $\phi$ and $\chi$:
\begin{subequations}
\label{eq:individual_eos}
\begin{align}
   w_{\phi}
   &\equiv
   \frac{P_{\phi}}{\rho_{\phi}}
   =
   \frac{\dot{\phi}^2-2V}{\dot{\phi}^2+2V}
   ,
   \\
   w_{\chi}
   &\equiv
   \frac{P_{\chi}}{\rho_{\chi}}
   =
   1
   .
\end{align}
\end{subequations}
From the Friedmann equations \eqref{eq:harm_ghost}, one can determine that
the energy density of the field $f$ scales as $\rho_f\propto a^{-3(1+w_f)}$.
%=============================================================================
\subsection{Initial conditions}\label{sec:cosmo_id}
For our initial conditions, we first set the free initial data
by superimposing the homogeneous initial conditions for
the cosmological scalar fields and metric with the
metric of a (rotating) black hole spacetime.
We then solve the constraint equations for the full
metric using a conformal thin
sandwich solver \cite{East:2012zn} 
(see appendix \ref{sec:numerical_methodology} for more details).

For the cosmological free initial data,
we consider an initially contracting FLRW universe dominated by the 
canonical scalar field $\phi$ (that is, with the
initial condition $\rho_{\chi}\ll\rho_{\phi}$). 
In this limit, with the potential $V=V_0e^{-c\phi}$, $\phi$
can obey a scaling solution such that the effective equation of state
is roughly constant and equal to
$w_{\phi}=\frac{c^2}{3\sqrt{16 \pi}}-1$ 
\cite{Allen:2004vz} (see more generally 
\cite{PhysRevD.32.1316,HALLIWELL1987341,Burd:1988ss}). 
For $c>\sqrt{96 \pi}$ and $V_0 <0$, 
the scaling solution in a contracting universe is \emph{ekpyrotic}:
the contracting solution is a dynamical attractor,
and density perturbations are smoothed out in each Hubble 
patch during contraction \cite{Garfinkle:2008ei, Khoury:2001wf,Erickson:2003zm,
Ijjas:2020dws,Cook:2020oaj,Ijjas:2021wml}.
In this limit $w_{\phi}\geq1=w_{\chi}$, so if $\rho_{\phi}>\rho_{\chi}$
initially during contraction, it remains so for all remaining time
(recall $\rho_f\propto a^{-3(1+w_f)}$), and there cannot be a bounce. 
We instead consider the scenario where 
$c<\sqrt{96 \pi}$ and $V_0 >0$ so that $w_{\phi} < 1 = w_{\chi}$,
which is required in order to obtain a nonsingular bounce with the
massless ghost field we consider \cite{Allen:2004vz,Xue:2013bva}.
As a result, the asymptotic, contracting, solution is \emph{not} an attractor
and the initial condition must be fine-tuned in order to keep $w_{\phi}$ 
constant during the contracting phase. 
We justify this by noting that our main goal is to just explore the bouncing
phase, and not to give a completely realistic description
of a bouncing cosmology. 

Setting $c<\sqrt{96 \pi}$ implies $w_{\chi}>w_{\phi}$, 
so the negative energy density of
$\chi$---which we choose to be initially negligible---grows faster than
the positive energy density of the
canonical field during the contraction.
Because of this, the total scalar field energy density 
$\rho_{\phi}+\rho_{\chi}$ eventually goes through zero,
and the sign of $\dot{a}$ switches from being negative to being positive. 
At this point, the universe goes from contraction to expansion.
From the Friedmann equations \eqref{eq:harm_ghost}, 
we see that once expansion has
begun, the ghost field energy quickly 
diminishes and becomes negligible again compared to 
the energy density of $\phi$ \cite{Allen:2004vz,Xue:2013bva,Xue:2013iqy}.

In~\eqref{eq:cosmo_harm_id}, we present our choice of asymptotic FLRW initial data, 
which, as discussed above, is fine-tuned to allow for the asymptotic
cosmological value of $w_{\phi}$ to
remain roughly constant during contraction up until the bouncing phase.
The initial values for $\phi$, $\phi'$, $\chi$, $\chi'$, $a$, and $a'$ are:
\begin{subequations}\label{eq:cosmo_harm_id}
\begin{align}
   \phi(0)\equiv \phi_0=0, \quad \dot{\phi}_0 =-{a_0}^3
\sqrt{\frac{32 \pi c^2 V_0}{96 \pi-c^2}} 
   ,\\
   \chi(0)\equiv \chi_0=0, \quad \dot{\chi}_0 ={a_0}^3
\sqrt{\frac{12 V_0}{(96 \pi-c^2)\eta_0}} 
   ,\\
   a(0)\equiv a_0=1, \quad \dot{a}_0 =-{a_0}^4
\sqrt{\frac{2 V_0(\eta_0-1)}{(96 \pi-c^2)\eta_0}} 
   .
\end{align}
\end{subequations}
Here $\eta_0\equiv \eta(0)$ is the initial value of the ratio between
the energy densities of the two scalar fields 
\begin{align}\label{eq:flrw_ratio}
   \eta \equiv \left|\frac{\rho_\phi}{\rho_\chi}\right|
   .
\end{align}
We compute $\rho_{\phi},\rho_{\chi}$ in the code using
formulas \eqref{eq:stress_energy} and \eqref{eq:energy_density}.

In a similar fashion to \cite{Xue:2013bva}, we choose
$c=\sqrt{48 \pi}$ so that $\phi$ initially behaves like matter with $w_{\phi}=0$.
Such a matter-like contracting phase can generate scale invariant 
adiabatic perturbations
that would seed structure formation in the early expansion phase.

%=============================================================================
\section{Overview of numerical method and diagnostics}\label{sec:diagnostics}

We evolve the system \eqref{eq:equations_of_motion} 
nonlinearly using the harmonic formulation, 
and work with an axisymmetric spacetime.
We spatially compactify our numerical domain, and evolve the boundary using
the homogeneous FLRW equations of motion \eqref{eq:harm_ghost}.
See appendix~\ref{sec:numerical_methodology} for a more thorough discussion 
on our numerical methods.

In order to characterize our results, we make use of 
several diagnostic quantities.
We define the following stress-energy tensors
\begin{subequations}\label{eq:stress_energy}
\begin{align}
   {T_{\mu\nu}}^{(\phi)} 
   &= 
   2 \nabla_{\mu} \phi \nabla_{\nu} \phi 
   -
   g_{\mu\nu} \left(\nabla^{\alpha}\phi \nabla_{\alpha} \phi + 2 V \right)
   ,\\
   {T_{\mu\nu}}^{(\chi)} 
   &=  
   - 2 \nabla_{\mu} \chi \nabla_{\nu} \chi 
   + g_{\mu\nu}  \nabla^{\alpha}\chi \nabla_{\alpha} \chi 
   ,
\end{align}
\end{subequations}
so that the Einstein equations read
\begin{align}
   R_{\mu\nu}
   -
   \frac{1}{2}g_{\mu\nu}R
   =
   8\pi \left(
      T_{\mu\nu}^{(\phi)}
      +
      T_{\mu\nu}^{(\chi)}
   \right)
   .
\end{align}
From ${T_{\mu\nu}}^{(\phi)}$ and ${T_{\mu\nu}}^{(\chi)}$ 
we define the corresponding energy densities
\begin{equation}
   \label{eq:energy_density}
   \rho = n^{\mu} n^{\nu} T_{\mu\nu}
\end{equation} where $n^{\mu}$ is the time-like unit normal vector 
to hypersurfaces of constant time. 
We additionally compute the local expansion rate
\begin{equation}\label{eq:Hubble}
   H_K \equiv -\frac{K}{3}
   ,
\end{equation}
where $K$ is the trace of the extrinsic curvature on each
constant $t$ time slice (as in, e.g., \cite{East:2015ggf,East:2016anr}).
We note that $H_K$ asymptotes to $H$ at the boundary of our domain
\begin{align}
   H
   =
   \lim_{r\to\infty}H_K
   ,
\end{align}
where $r$ is the proper circumferential radius 
(see equation~\eqref{eq:proper_r}).
We define an effective scale factor on each time slice
\begin{equation}\label{eq:aeff}
	a_{\rm eff} (t,\vec{x}) \equiv |\gamma_3|^{1/6}
\end{equation}
where $\gamma_3$ is the determinant of the (three-dimensional)
metric intrinsic to each constant time hypersurface.
We are mainly interested in computing 
\eqref{eq:energy_density} to 
\eqref{eq:aeff} on the black hole surface, and at different coordinate radii 
far away from the black hole. 
For non-rotating black holes, we track their values as a function of the distance
from the center of the black hole. 
We compare the values to their homogeneous 
counterpart given by \eqref{eq:flrw_energies} 
and \eqref{eq:flrw_hamiltonian_constraint}. 
In axisymmetric spacetimes, the coordinate radius on the equator $r_{\rm co}$ 
is related to the proper circumferential radius $r$ through the relation
\begin{equation}\label{eq:proper_r}
   r \left(t, \theta=\frac{\pi}{2}\right) 
   = 
   \sqrt{\gamma_{zz}\left(t, \theta=\frac{\pi}{2}\right)} 
   \ 
   r_{\rm co}(t)
   .
\end{equation}
where $\gamma_{zz}$ is the value of the spatial metric along the symmetry axis.
In spherical symmetry, Eq.~\eqref{eq:proper_r} reduces to the areal radius. 
To characterize the boundaries of black holes in our dynamical setting, 
we will consider
two surfaces: event horizons and apparent horizons.
The black hole event horizon is the boundary behind which null rays no
longer escape to the asymptotic region.
We compute its approximate location 
by integrating 
null surfaces backwards in time 
\cite{Anninos:1994ay,Libson:1994dk,Thornburg:2006zb}
(we restrict this to spherically symmetric cases, where it
is sufficient to consider spherical null surfaces).
We define the apparent horizon of the black hole, on the other hand,
on each time slice, as the outermost marginally outer
trapped surface, i.e. the surface for which the outgoing null expansion 
$\theta_{(l)}$ vanishes and the inward null expansion $\theta_{(n)}$ 
is negative and such that $\theta_{(l)} > 0$ immediately outside the black hole
(and $\theta_{(l)} < 0$ immediately inside)
\footnote{In our particular setup, we define the outgoing (ingoing)
direction as the direction pointing from the origin (asymptotically FLRW region)
to the asymptotically FLRW region (the origin)}.

In analogy to black hole apparent horizons, 
we will also use marginally trapped surfaces to define the location
of the \emph{cosmological apparent horizon}. 
We will refer to this simply as the cosmological horizon, but we 
note that this is not to be confused with the \emph{event horizon}
or the \emph{particle horizon} commonly used in cosmology.
During the contracting phase,
the cosmological horizon is defined as the surface for which the outgoing
null expansion $\theta_{(l)}$ vanishes, and the inward null expansion
$\theta_{(n)}$ is negative, but $\theta_{(l)} > 0$ immediately inside the
cosmological horizon.  During the expanding phase, the cosmological horizon is
defined as the surface for which the ingoing null expansion $\theta_{(n)}$
vanishes and the outward null expansion $\theta_{(l)}$ is positive and such
that $\theta_{(n)} > 0$ outside the cosmological horizon.  

In a homogeneous spacetime, the cosmological apparent horizon is simply
the sphere with coordinate radius equal to the 
comoving Hubble radius, $R_H = (aH)^{-1}$, yielding an area 
$A_{\rm C} = 4 \pi a^2 {R_H}^2 = 4 \pi H^{-2}$.
For our black hole spacetimes, we will always take the cosmological horizon
to be centered on the black hole. This is because we are interested in the dynamics
in the vicinity of the black hole, and it is this surface that is the most relevant
to understanding the behavior of the black hole horizon.
See appendix~\ref{sec:numerical_methodology} for more details on our numerical
implementation and appendix~\ref{sec:DH} for more details on the various
definitions of horizons we use. 

From the area of the black hole apparent horizon $A_B$, we define an areal mass 
$M_{\rm A} \equiv \sqrt{A_B/(16 \pi)}$. The spacetime we study here violates the NCC,
and thus we expect to find instances where $M_A$ decreases.
Similarly, the second law of black hole thermodynamics states that so long as
the NCC is satisfied, the area of a black hole event horizon must increase
into the future \cite{PhysRevLett.26.1344}.
This can be extended to the cosmological setting assuming
that the universe does not again collapse, and 
a notion of infinity can be defined~\cite{Carr:1974nx}.
However, here
we are evolving a black hole in a spacetime that violates the NCC,
and find that the event horizon does decrease in area.

The cosmological and black hole apparent horizons that we find on each
time slice can also be thought of as foliations of three dimensional 
surfaces called  
holographic screens 
\cite{Bousso:1999cb,Bousso:2015mqa,Bousso:2015qqa} or
Marginally Trapped Tubes (MTTs) \cite{Ashtekar:2005ez}
in general, and dynamical horizons
\cite{Hayward:1993mw,Ashtekar:2002ag,Ashtekar:2003hk}
if they obey certain extra conditions
(we review the definitions of these concepts in appendix~\ref{sec:DH}).
Though one can formulate area laws for these surfaces, in spherical symmetry
they do not place any constraints on whether the area increases to the future.
We keep track of the MTTs corresponding to the cosmological and black hole
apparent horizons, and in particular, compute when they are spacelike or
timelike in nature. 

For the black holes, we compute the 
equatorial circumference of the horizons 
$c_{\rm eq}$, and define their corresponding equatorial radii 
$r_{\rm eq} = c_{\rm eq}/ 2\pi$,
which in the case of spherical symmetry is also equal to the areal radius, 
$r_A =\sqrt{\frac{A_B}{4\pi}}$.
When studying rotating black holes we can also associate an angular momentum to 
the apparent horizons
\begin{equation}\label{eq:J_ah}
J_{\rm AH} \equiv \frac{1}{8\pi} \int {\hat{\phi}}_i K^{ij} dA_j ,
\end{equation}
where ${\hat{\phi}}_i$ is the axisymmetric Killing vector, and, using the 
Christodoulou formula, we can define a mass
\begin{equation}\label{eq:mass_ah}
M_{\rm AH} \equiv \left({M_A}^2 +\frac{J_{\rm AH}^2}{4 {M_A}^2}\right)^{1/2}.
\end{equation}
Since the scalar fields do not carry any angular momentum in axisymmetry,
the total angular momentum 
of the black hole remains constant throughout the evolution of our
spacetime. Thus, we will only be interested in 
the total mass and circumferential radius of the black hole.

%=============================================================================
\section{Results\label{sec:numerical_results}}

%-----------------------------------------------------------------------------
We begin by studying the evolution of non-spinning black holes in 
an asymptotically bouncing universe 
(section~\ref{sec:low_mass}-\ref{sec:extrapolate}) 
using the method described in section~\ref{sec:diagnostics}.  
Though we do not explicitly enforce spherical symmetry, we find
no evidence of any instabilities that break that symmetry if our
initial data respects it.
We consider non-spinning black holes in section~\ref{sec:spinning_bh}.

We find that the qualitative behavior of our solutions
can be divided into two regimes, which can be distinguished by the
ratio of the areal radius of the initial
black hole horizon, $r_{\rm{BH},0}$
and the minimum size
of the Hubble radius of the background cosmology
$R_{\rm{H},\rm{min}} \equiv \min_t \rvert 1/H \rvert = -1/H_{\rm min}$
(where $H_{\rm min}<0$ is the maximum contraction rate). 
When $R_{\rm{H},\rm{min}}/r_{\rm{BH},0}\gtrsim3.5$, 
the black holes pass through the bounce freely. 
When $R_{\rm{H},\rm{min}}/r_{\rm{BH},0}<3.5$, we find that 
the locally defined cosmological and black hole apparent horizons merge, 
and cease to exist for a period of time during the contracting 
and bouncing phase.
We note that the horizons merger at $R_{\rm{H},\rm{min}}/r_{\rm{BH},0}>1$,
as the black hole grows in size during the contraction phase
(see figures \ref{fig:m_AH},\ref{fig:2m_AH}; 
we will discuss this more in the following
subsections).

For every initial data setup we considered, 
we find that the black hole continues to exist after the bounce phase ends:
the late-time evolution always
consists of a black hole in an expanding universe
with the ghost field energy density decreasing
at a faster rate than the canonical scalar field energy. 
Moreover, we find that the late time black hole mass remains similar to the
initial black hole mass, regardless of the ratio of the initial black
hole radius and minimum Hubble patch radius.
In the following sections, we quantify these observations and
extrapolate our findings to the regime where the Hubble radius 
shrinks to a much smaller size compared to the radius of the black hole.
%-----------------------------------------------------------------------------
\subsection{Small black hole regime\label{sec:low_mass}}
We first consider solutions where $R_{\rm{H},\rm{min}}/r_{\rm{BH},0}\gtrsim3.5$
(see above for definitions).
In figure~\ref{fig:m_rho}, we show the Hubble parameter (left panel) computed from
\eqref{eq:Hubble} and the ratio of scalar fields (right panel) computed from
\eqref{eq:flrw_ratio},\eqref{eq:stress_energy} and \eqref{eq:energy_density} as a function of harmonic time for different coordinate radii.
We also plot the value these quantities take
at spatial infinity, where we assume homogeneous FLRW boundary conditions
(see section~\ref{sec:equations_of_motion}).
While the bounce seems to be pushed to slightly earlier harmonic
times when the black hole is present, 
most of the local cosmological evolution remains unaffected by the presence of
the black hole and follows the same qualitative evolution 
as the background cosmology (section~\ref{sec:cosmo_id}).
To determine how the cosmology is affected in a region close to the black hole, 
in figure~\ref{fig:spatial_m}
we plot the spatial dependence of
$\eta$ and $H_K/|H_{\rm min}|$ as a function of distance
again along the equator at different times. 
Although the local expansion 
rate and the ratio of the energy densities can differ from their background values 
by up to $15-60 \%$ and $9-16 \%$ respectively, 
beyond $r \sim 10-25 r_{\rm{BH},0}$ both quantities quickly asymptote 
to their respective background values.
Note that the coordinate radius differs from the proper radius by the local 
scale factor; see eq.~\eqref{eq:proper_r}.
The effective scale factor computed from \eqref{eq:aeff} 
at different coordinate radii 
is plotted in figure~\ref{fig:aeff} (see appendix \ref{sec:numerical_methodology}). 
Again we find that
far enough from the black hole,
the value of 
scale factor remains largely unaffected by the presence of the black hole. 
We caution that these quantities will also be subject to gauge effects---in
particular from our choice of the lapse function
(see section~\ref{sec:numerical_methodology}).
As we describe below, towards the end of the simulations we find strong variation
in the rate of which time advances at different spatial points.

\begin{figure}[h]
\centering
\includegraphics[width=.495\textwidth]{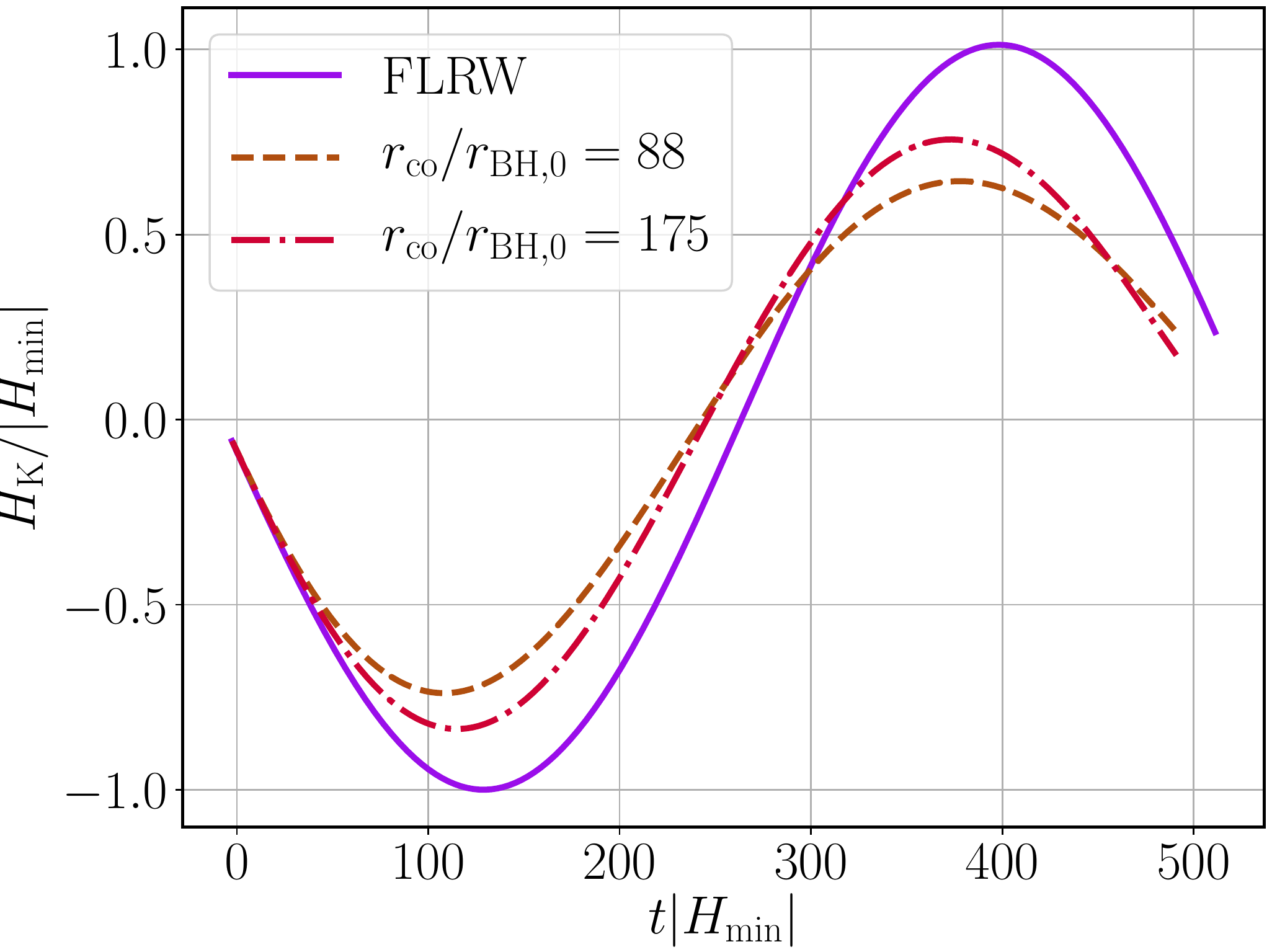}
\includegraphics[width=.495\textwidth]{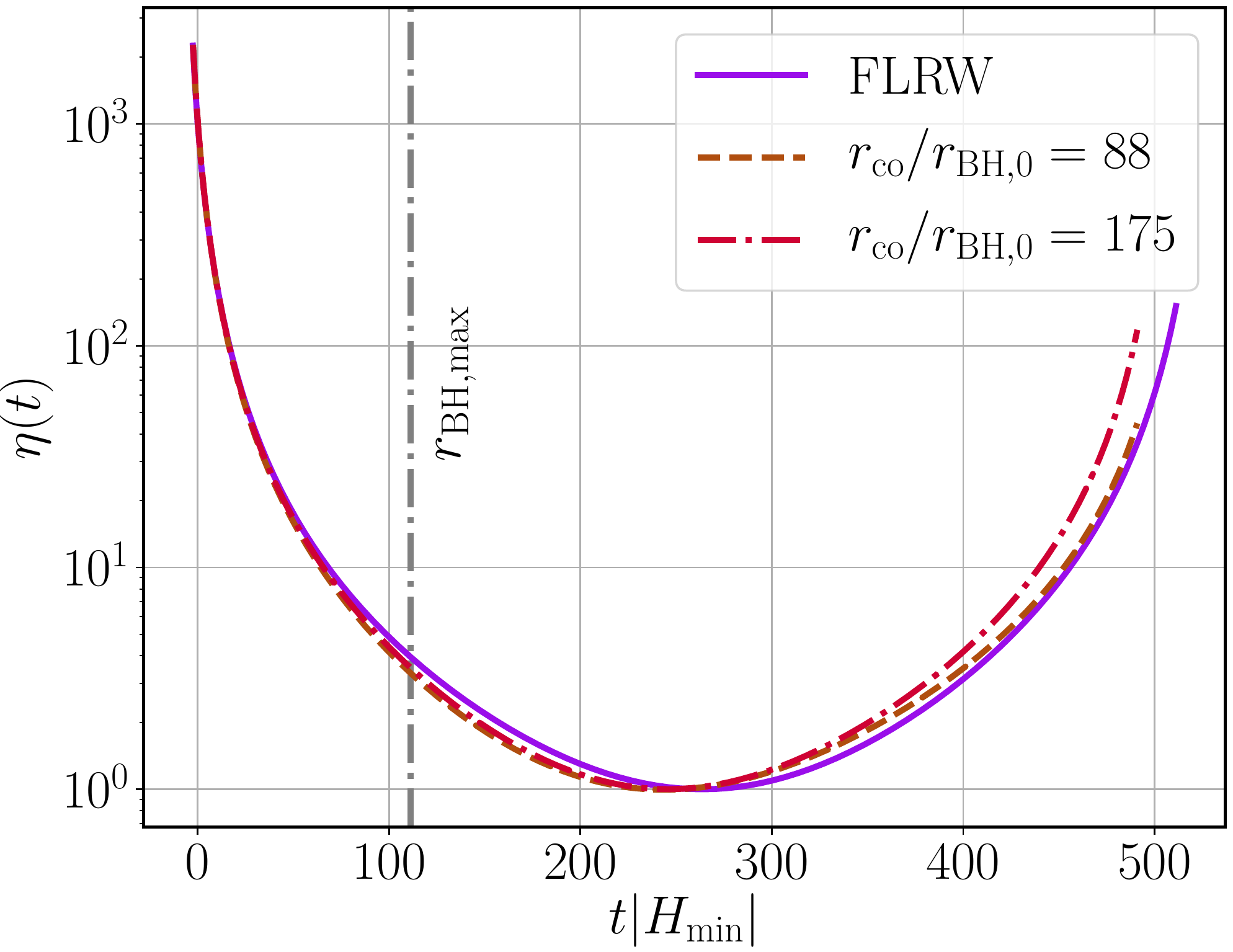}
\caption{\label{fig:m_rho} 
   The expansion rate $H_K/ |H_{\rm min}| $ computed from 
   \eqref{eq:Hubble} (left) and the ratio of the matter to ghost 
   field $\eta(t)$ given by \eqref{eq:flrw_ratio} (right) for a black hole with 
   initial mass such that the Hubble radius of the background cosmology
   $R_{\rm{H}} \equiv |H^{-1}|$ shrinks from an initial value of 
   $R_{\rm{H},0} = 75 r_{\rm BH,0}$ to $4.34 r_{\rm BH,0}$
   (here $r_{\rm BH,0}$ is the initial black hole radius). 
   The solid line shows the corresponding background 
   solution, and the dashed and dash-dotted lines show the values at 
   different coordinate radii. 
   The vertical grey line is the time at which the 
   black hole reaches its maximum areal radius as measured by the apparent horizon.
   Notice that the black hole reaches its maximum size slightly
   before the universe at large scales bounces, as the ghost field
   begins to dominate at an earlier time the closer one gets to the
   black hole horizon.
   The slight difference in the maximum absolute value of
   the FLRW value of
   $H_{\rm{K}}/|H_{\rm{min}}|$ at $t|H_{\rm{min}}|\sim 120,400$ is
   due to numerical error in our integration.
}
\end{figure}

\begin{figure}[h]
\centering
\includegraphics[width=\textwidth]{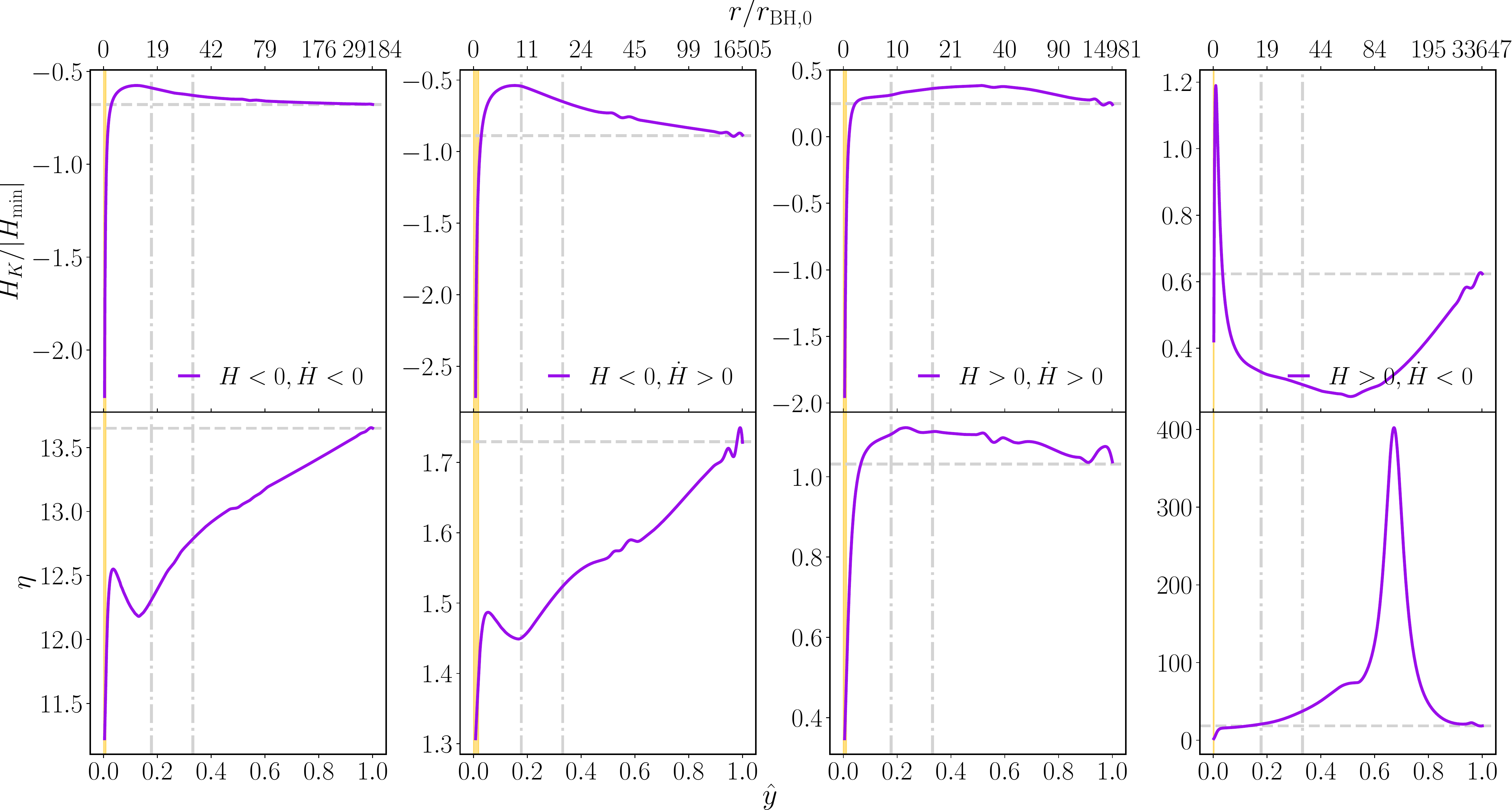}
\caption{\label{fig:spatial_m} 
The expansion rate $H_K/ |H_{\rm min}| $ (top) and the ratio of the energy
densities of the matter to ghost field $\eta(t)$ given by \eqref{eq:flrw_ratio}
(bottom) for the black hole considered in figure~\ref{fig:m_rho} as a function
of the compactified coordinate radius, $\hat{y}$
(see equation~\eqref{eq:compactified_coordinate_relation}), at different
times during the evolution. Note that $\hat{y}$ lies along the ``equator''
of the black holes in our simulations.
Also shown on the top axis is the proper radius of
the spacetime computed from \eqref{eq:proper_r}. The dashed horizontal grey
lines indicate the corresponding background values at spatial infinity, the
vertical dash-dotted lines correspond to the coordinate radii shown in
figure~\ref{fig:m_rho}, and the shaded region represents the black hole.
}
\end{figure}

\begin{figure}[h]
\centering
\includegraphics[width=.495\textwidth]{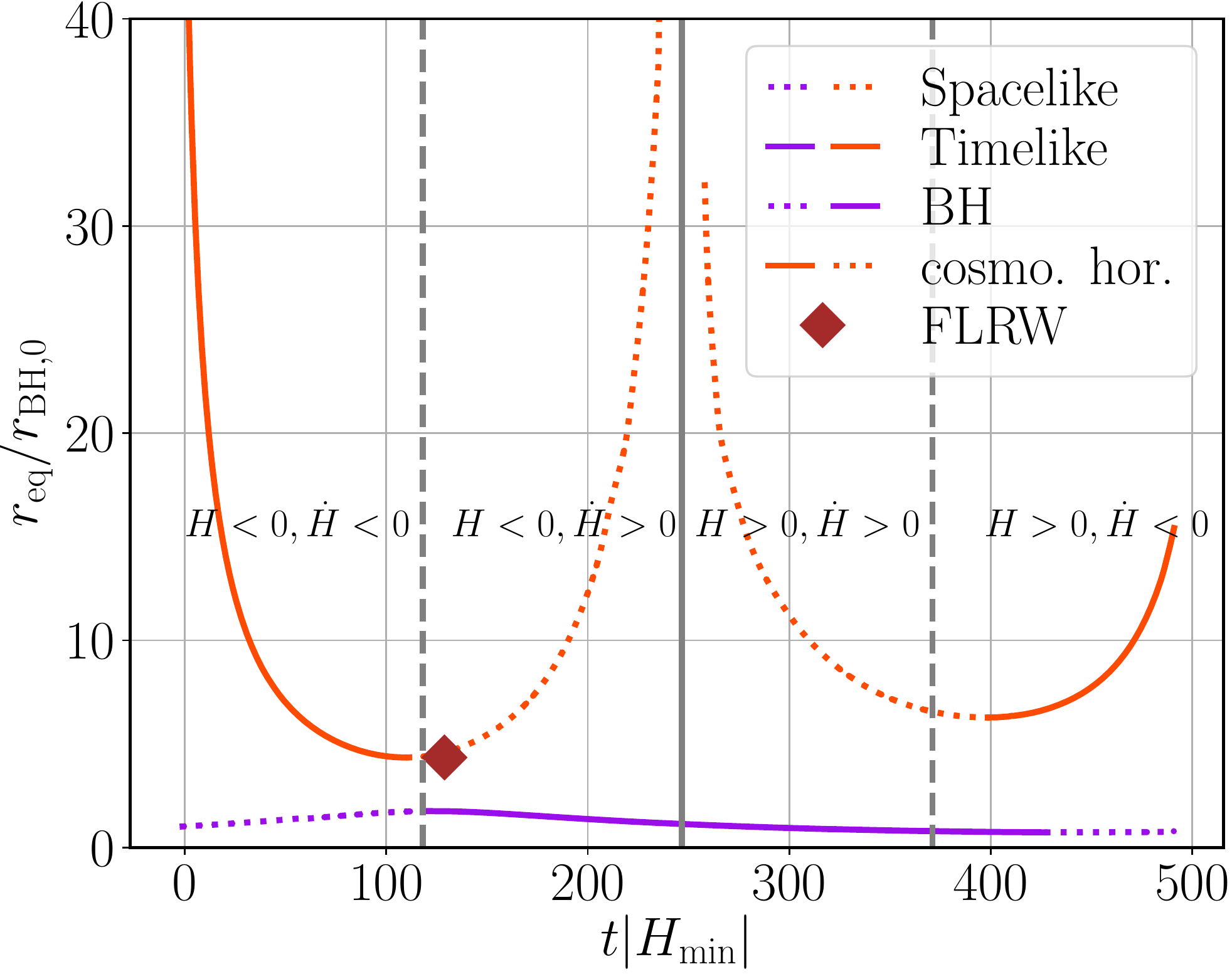}
\includegraphics[width=.495\textwidth]{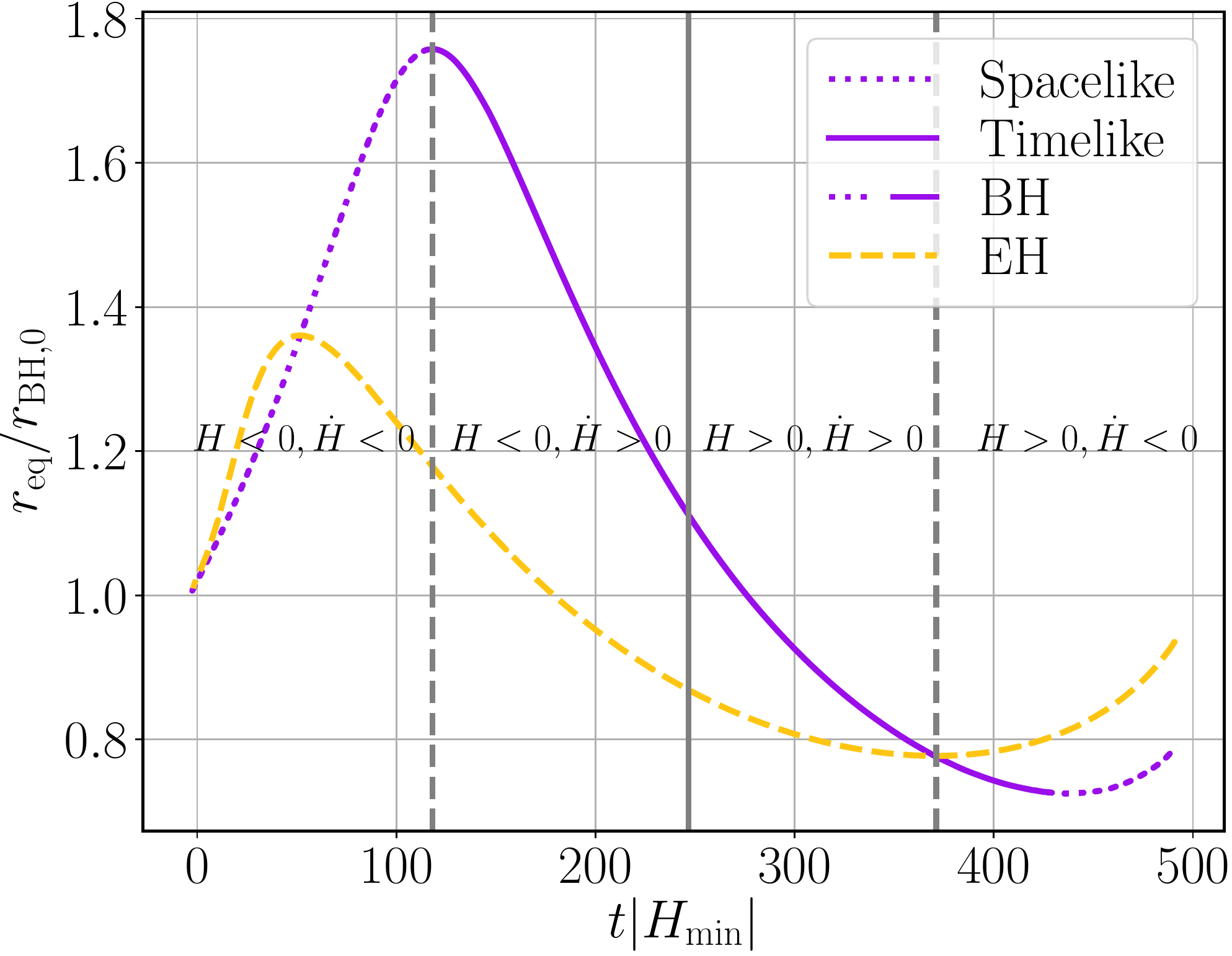}
\caption{\label{fig:m_AH} 
\textit{Left}: The areal radii of the black hole (purple) and 
   cosmological (orange) apparent horizons computed as described in 
   section~\ref{sec:diagnostics} for the black hole in figure~\ref{fig:m_rho}. 
   The line style reflects the signature of the marginally trapped tube 
   or holographic screen (solid is timelike, dashed is spacelike). 
   The diamond indicates the time and value 
   to which the Hubble radius would shrink in the absence of a black hole.
   \textit{Right}: The apparent horizon of the black hole (purple) and 
   the corresponding event horizon (yellow dashed). 
   The vertical solid line indicates the bounce, while the region between 
   the dashed lines is the bouncing phase (where the NCC is violated).
}
\end{figure}

\newpage
We next present several results regarding 
the behavior of the area of the black hole,
as measured by either the event or apparent horizon. 
Naively, one expects the accretion of the canonical/ghost field to result 
in an increase/decrease in mass of the black hole \cite{Babichev:2004yx}.
That being said, it is less clear how a black hole embedded in a cosmology
driven by a canonical/ghost field may behave \cite{Gao:2008jv,Gao:2011tq}.

Figure \ref{fig:m_AH} depicts the evolution of the 
black hole's areal radius.
We find that during the contracting phase prior to the
bouncing/NCC violation phase,
the canonical scalar field energy density exceeds that of the ghost field; 
see figure~\ref{fig:m_rho} and the solid purple curve in figure~\ref{fig:spatial_m}.
The black hole's proper area increases during this time;
(first region in figure~\ref{fig:m_AH} where $H<0,\dot{H}<0$). 
Once the bouncing phase starts
($t |H_{\rm min}| \sim 120$ in figure~\ref{fig:m_rho}), 
the black hole starts to shrink 
as one may expect since 
the ghost field energy in this regime is comparable to the canonical 
scalar field energy density 
(second region where $H<0$ and $\dot{H}>0$ in figures \ref{fig:m_rho} and \ref{fig:spatial_m}). 
Near the end of the bouncing phase the universe is expanding (region where $H>0$ and $\dot{H}<0$),
yet the black hole's size is still shrinking in this region, as the
ghost field energy density still dominates over the canonical
scalar field energy density in the region near the black hole
(in other words, $\eta<1$ in the region close to the black hole, 
see figure \ref{fig:spatial_m}), although at an increasingly slower 
rate as the ghost field energy 
density quickly diminishes in time. 
After the end of the bouncing phase,
the universe continues to expand, 
the ghost field decays to dynamically irrelevant values,
and the black hole begins growing in size 
(fourth region where $H>0$ and $\dot{H}<0$ and the dotted purple curve). 

The left panel of figure~\ref{fig:m_AH} also shows 
the areal radius of the cosmological horizon.
We see that during the contracting phase, the cosmological horizon
shrinks from $r_{\rm{C},0}= 75 r_{\rm{BH},0}$ to
a minimum radius of $r_{\rm{C},\rm{min}}= 4.34 r_{\rm{BH},0}$ at 
$t \sim 50 r_{\rm{BH},0}$. 
This is similar to the value the Hubble radius ($R_{\rm{H}} \equiv |1/H|$),
would shrink to in the absence of a black hole. 
This value is indicated by the diamond in figure~\ref{fig:m_AH}.
From this we conclude that---at least in this regime---the presence
of the black hole does not qualitatively change the dynamics of the spacetime.
Past this point of closest encounter,
the cosmological horizon tends to $r_C \to + \infty$
which defines the location of the bounce ($\lim_{H\to0}1/H=\infty$). 
Once the universe switches from contraction to expansion, 
the cosmological horizon is defined as the location where the ingoing null 
expansion vanishes and outgoing null expansion is positive. 
After the bounce, the cosmological horizon
at first shrinks to a minimum size before re-expanding to $+ \infty$. 
We note that the areal radius of the cosmological
horizon is no longer symmetric about the bounce once a black hole is present. 

We also compute the signature of the 
MTTs associated with the horizons (see appendix~\ref{sec:DH} for definitions), 
which we plot in figure~\ref{fig:m_AH}. 

First we study the properties of the black hole MTT in more detail. 
Using the terminology of appendix~\ref{sec:DH}, the
black hole is a future marginally trapped tube foliated by 
future marginally outer trapped surfaces (alternatively called a
future holographic screen). 
The area law of dynamical horizons states that if the MTT is spacelike 
(i.e. if it is a dynamical horizon), 
then the area of the black hole should increase in the
outward radial direction, while if the MTT is timelike
(i.e. we have a timelike membrane with $\Theta_{(n)} <0$), 
then the area should increase into the past.
Looking at figure~\ref{fig:m_AH} 
we find that (as expected)
these laws are obeyed at all times, even during the bouncing phase.

We next look at the cosmological horizon.
We consider the contracting and expanding phases separately. 
During the contracting phase, the cosmological horizon is a MTT 
foliated by future marginally inner trapped surfaces
(alternatively, it is a future holographic screen).
From the area law of future holographic screens 
\cite{Bousso:2015mqa,Bousso:2015qqa},
we expect the cosmological horizon 
to obey the same area law as the black hole during the contracting phase.
Our findings agree with this expectation:
we find that the cosmological horizon is timelike when it
decreases in time and spacelike when it increases in the outward direction.
During the expanding phase, however, the cosmological horizon ceases to be a MTT.
Instead, we find that it satisfies the definition of a past holographic screen
(as the ingoing null expansion now vanishes).
From~\cite{Bousso:2015mqa,Bousso:2015qqa}, 
we still expect its area to increase in the future on timelike portions 
and in the outward direction on spacelike portions. 
Again we find that this is satisfied at all times during the expanding phase. 

We conclude by looking at the event horizon
shown in the right panel of figure~\ref{fig:m_AH}. 
Our main finding here is that the event horizon no longer lies 
outside the apparent horizon at all times.
This is a result of the violation of the NCC \cite{hawking1973large}. 
Interestingly, this behavior
begins not during the bouncing phase of cosmological evolution
(between the two dashed grey lines)
when the NCC is violated, but \emph{before} 
the bouncing phase has begun.
This is because the event horizon is not a quasi-local quantity, so
it can ``anticipate'' the bouncing/NCC violation phase.
In general, we find that the event horizon always increase until it crosses the 
apparent horizon of the black hole, after which it decreases.
Once the bouncing phase ends, the event horizon crosses the apparent horizon
again, after which it starts increasing and remains larger than it for all future times.

We were not able to evolve the spacetime to arbitrarily large proper times.
We ascribe this to gauge artefacts which impede the stable numerical evolution
of the solution.
In particular, the lapse function appears to become distorted in
the spacetime region between the 
black hole and the asymptotically homogeneous regime, which
causes that interior region to advance in time
much faster compared to elsewhere in the simulation. 
(This is evident in the rightmost panels
of figures~\ref{fig:m_rho},~\ref{fig:2m_rho}, and~\ref{fig:aeff}.)
That being said, based on the simulations we have run, we conjecture that
the black hole asymptotes to close to its initial mass at $t\to\infty$
with no significant gain or loss of energy.
That is, the end state is described by a black hole embedded in 
an expanding matter like FLRW universe with 
a negligible amount of ghost field and matter energy density.
This is illustrated in figure~\ref{fig:0p5m_AH} of 
appendix~\ref{sec:numerical_methodology} 
where we consider a black hole with half the mass of the one depicted
in figure~\ref{fig:m_AH}, i.e. we consider a black hole such that
the ratio of the minimum Hubble radius to the initial radius of the black hole
is $R_{\rm{H},\rm{min}}/r_{\rm{BH},0} = 8.69$.
Figure~\ref{fig:0p5m_AH} shows that, overall, the black hole's size changes by a negligible amount.
In this particular case, the final size of the apparent horizon of 
the black hole is $\sim 6\%$ larger that its initial value, the small difference
being an artefact of the initial data.
More importantly, figure~\ref{fig:0p5m_AH} also shows that the event horizon asymptotes 
to the apparent horizon at late times.

%-----------------------------------------------------------------------------
\subsection{Large black hole regime\label{sec:high_mass}}

We next consider solutions where 
$R_{\rm{H},\rm{min}}/r_{\rm{BH},0}<3.5$. 
The nonlinear evolution of one particular case 
is shown in figure~\ref{fig:2m_rho}.
As is the case for the lower initial mass evolutions 
(figure~\ref{fig:m_rho}), we 
see that the cosmological evolution remains unaffected 
far away from the black hole.
The bounce is pushed to even earlier times, as one may expect
since a large black hole could presumably accelerate the rate
of cosmological contraction.
Figure~\ref{fig:spatial_2m} shows that in the region near the black hole apparent horizon, 
the local expansion rate and the ratio of the energy densities 
now differ from their background value 
by up to $15$--$75 \%$ and $13$--$60 \%$.
Beyond $r \sim 2$--$12 r_{\rm{BH},0}$, both quantities asymptote 
to their respective background values.

\begin{figure}[h]
\centering
\includegraphics[width=.495\textwidth]{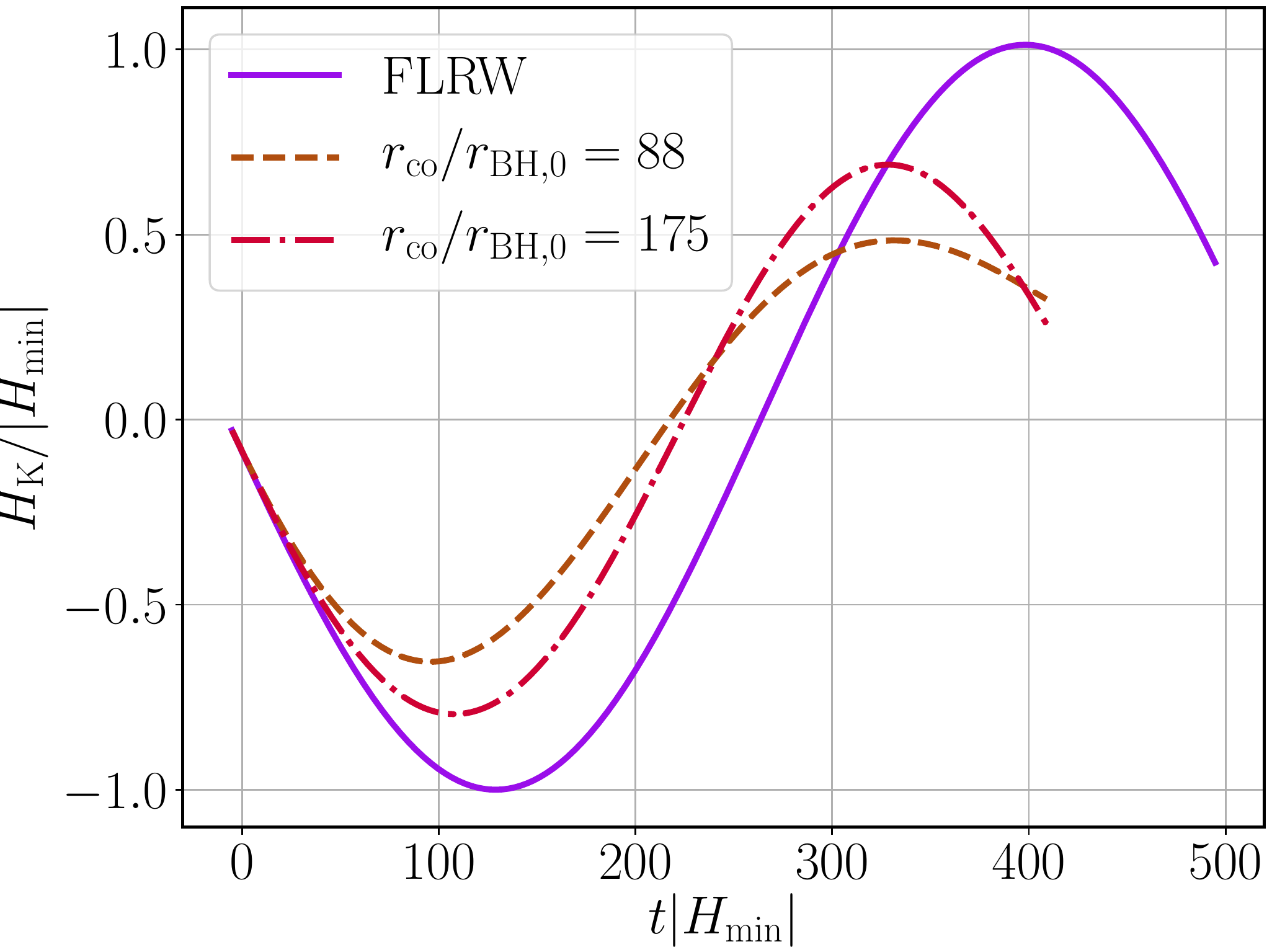}
\includegraphics[width=.495\textwidth]{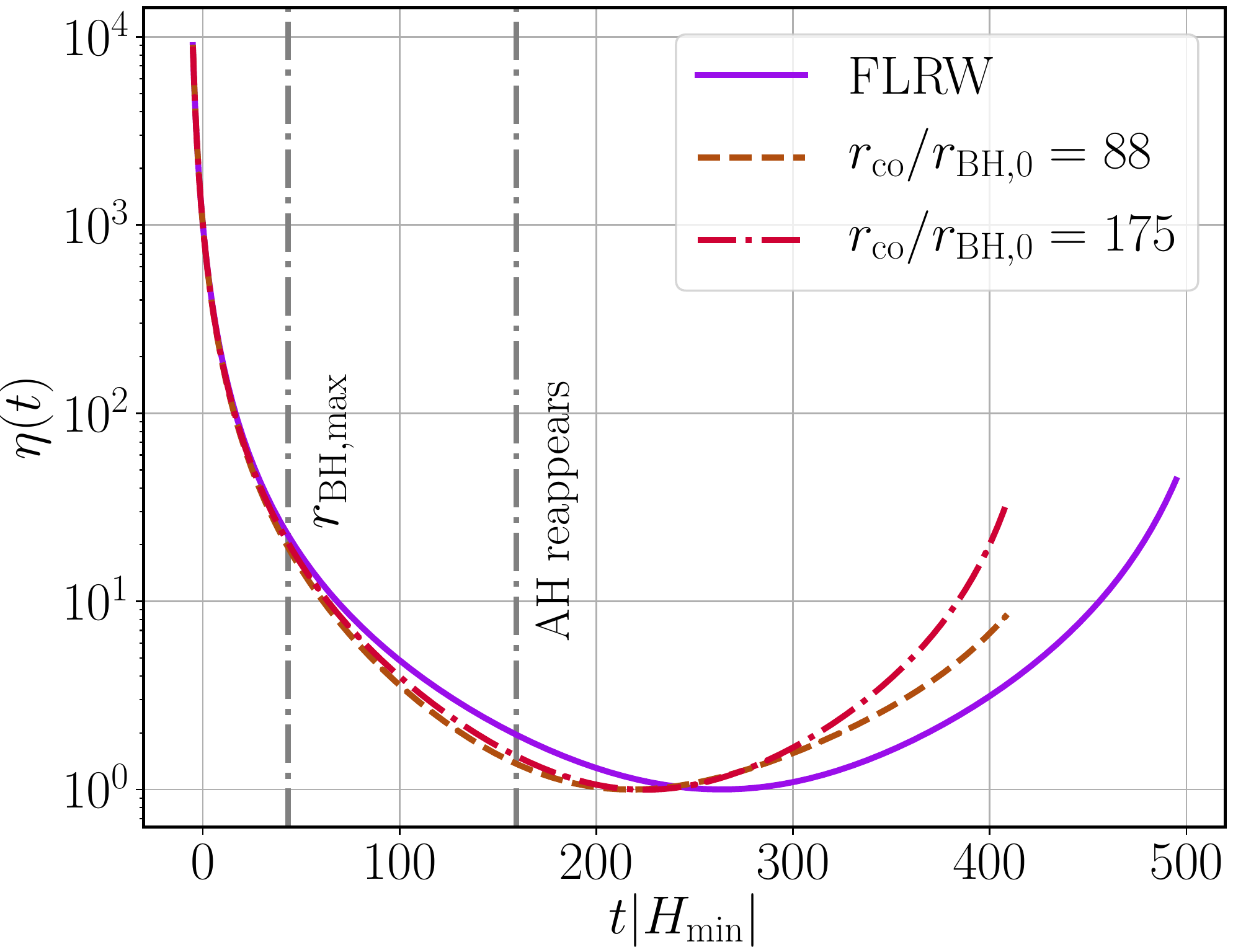}
\caption{\label{fig:2m_rho} 
   Same as figure~\ref{fig:m_rho}, but for a black hole with initial mass 
   such that the Hubble radius of the background cosmology
   $R_{\rm{H}} \equiv |H^{-1}|$ shrinks from an initial value of 
   $R_{\rm{H},0} = 75 r_{\rm BH,0}$ to $2.17 r_{\rm BH,0}$
    (here $r_{\rm BH,0}$ is the initial black hole radius).
}
\end{figure}

\begin{figure}[h]
\centering
\includegraphics[width=\textwidth]{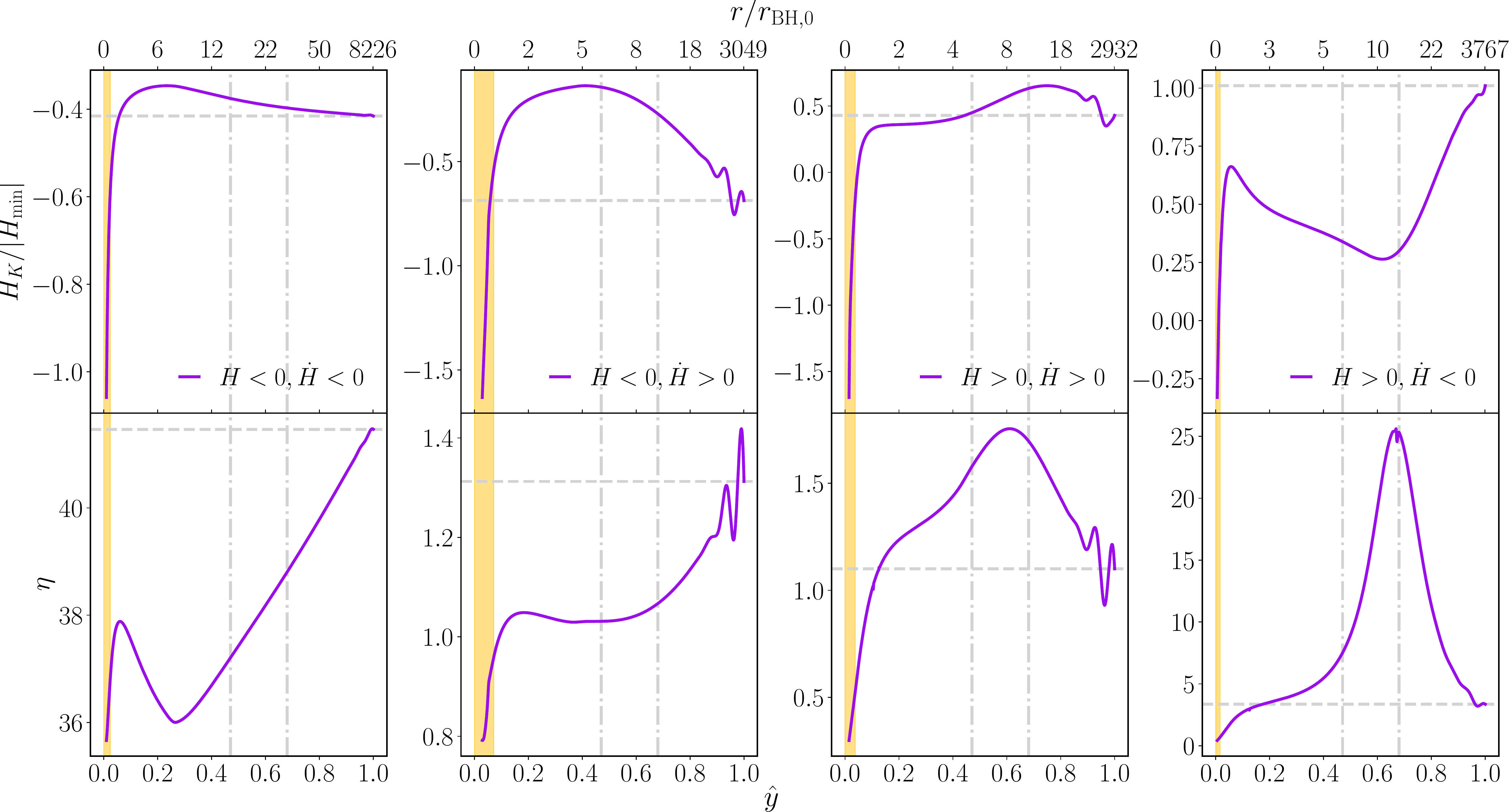}
\caption{\label{fig:spatial_2m} 
    Same as figure~\ref{fig:spatial_m}, but 
    for the case with $R_{\rm H,min}/r_{\rm BH,0}=2.17$ also shown 
    in figure~\ref{fig:2m_rho}.
}
\end{figure}

The behavior of the black hole and cosmological apparent horizons, 
which is shown in figure~\ref{fig:2m_AH}, 
is qualitatively different for the large black hole initial
data as compared to the small black hole initial data
($R_{\rm{H},\rm{min}}/r_{\rm{BH},0}\gtrsim3.5$). 
Similar to the cases studied in the section~\ref{sec:low_mass},
the cosmological horizon shrinks at first. 
Unlike those earlier cases though, it eventually merges
with the expanding black hole apparent horizon. 
Following the merger, the spacetime has no apparent horizons for
some time until they re-emerge.
After that, the cosmological and black hole apparent horizons follow 
a similar trajectory to the horizons studied in section~\ref{sec:low_mass}
during the cosmological expansion phase.

The merging of black hole and cosmological apparent horizons
has been observed in McVittie spacetimes 
\cite{Faraoni:2012gz,Faraoni:2014nba} 
(see also appendix~\ref{sec:mcvittie})
and can be interpreted the following way. 
As the apparent horizon of the black hole grows and the 
cosmological horizon shrinks during 
the contraction of the universe, 
we reach a point in time at which the black hole horizon 
coincides with the cosmological horizon.
At this point, one cannot distinguish between the black hole 
and the cosmological horizon (recall that during the contraction
the outward null expansion is negative outside of the cosmological horizon). 
A finite time later, before the bounce, but after the 
background Hubble radius reaches its minimum size, 
the effective Hubble radius has increased to a sufficiently large value
so that the black hole
solution again fits within the cosmological horizon.
At this point, the cosmological and black hole apparent horizons reappear.
We note that the black hole event horizon persists throughout the evolution of the
spacetime, so in this sense the black hole never disappears; see
figure~\ref{fig:2m_AH}.
We next investigate the physical properties of this process in more detail.  

\begin{figure}[h]
\centering
\includegraphics[width=.495\textwidth]{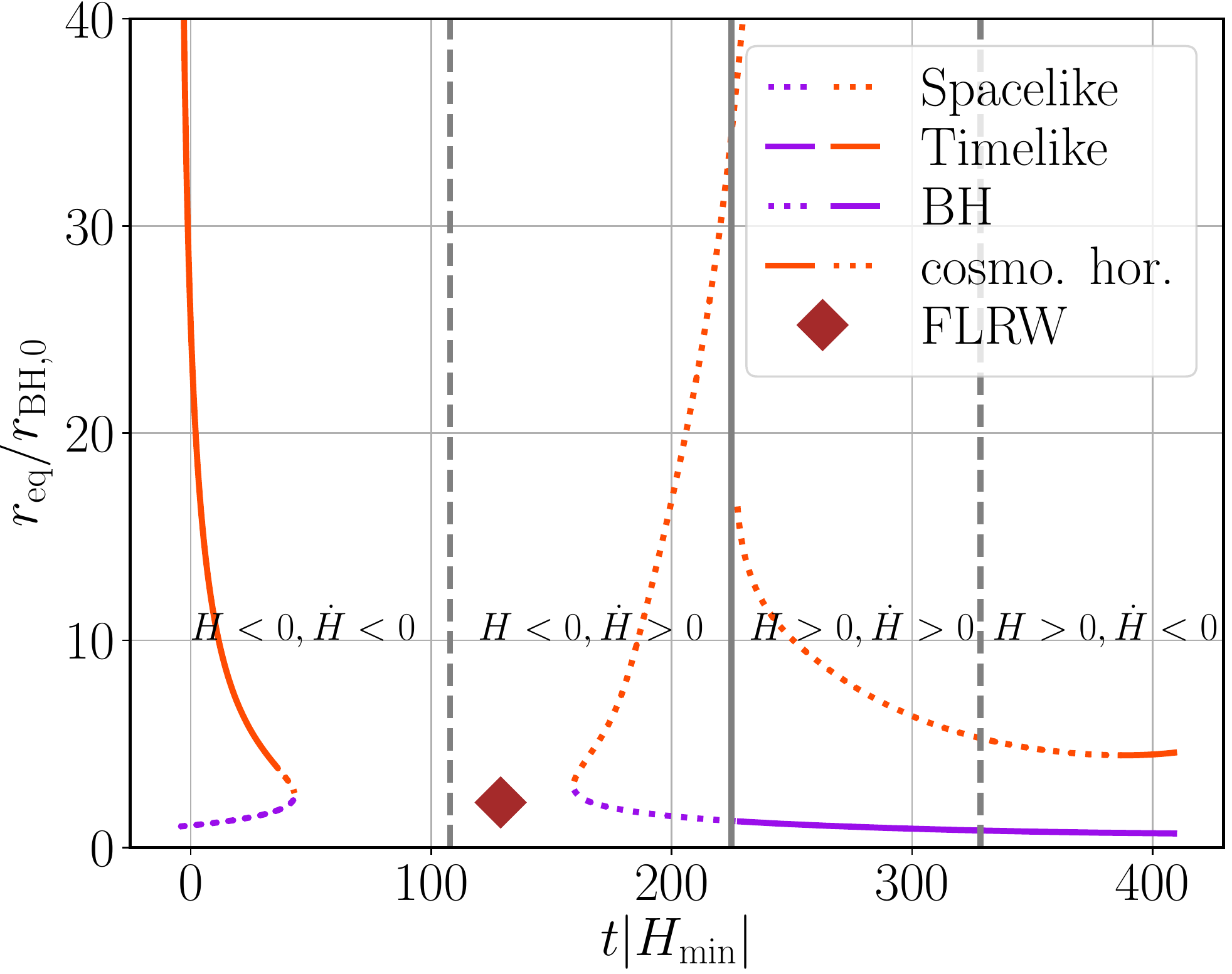}
\includegraphics[width=.495\textwidth]{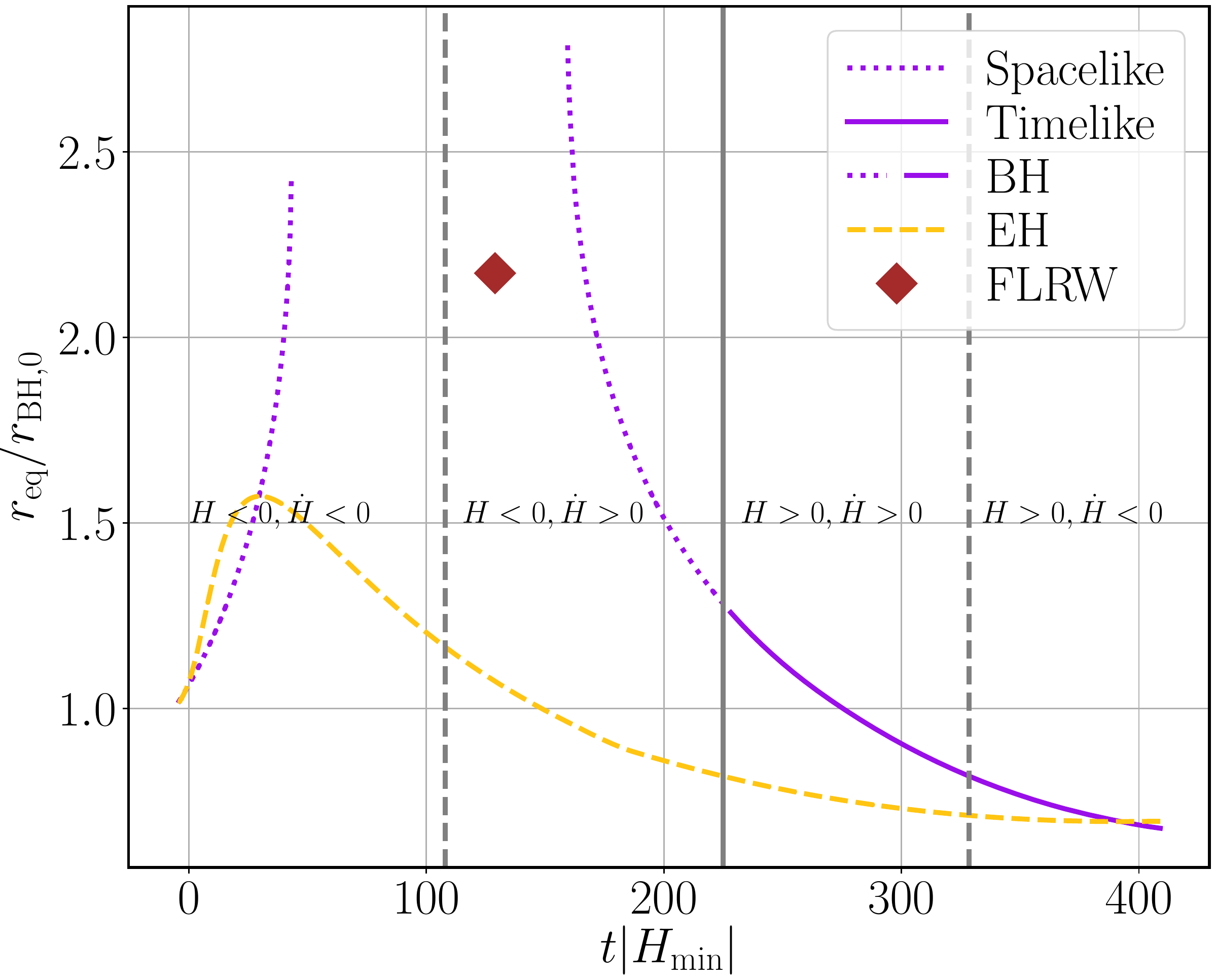}
\caption{\label{fig:2m_AH} 
   Same as figure~\ref{fig:m_AH} but for a black with initial 
   mass such that the Hubble radius of the background cosmology
$R_{\rm{H}} \equiv |H^{-1}|$ shrinks from an initial value of 
$R_{\rm{H},0} = 75 r_{\rm BH,0}$ to $2.17 r_{\rm BH,0}$
    (here $r_{\rm BH,0}$ is the initial black hole radius). 
    Notice that the location where $H=0$ 
    (that is, where the Hubble radius diverges) does
    not exactly coincide to where the cosmological horizon blows up,
    as the cosmological horizon is measured locally (in the interior
    of the computational domain), while $H=0$ is determined
    by the asymptotic cosmological evolution.
    For more discussion on how we define the cosmological horizon,
    see section~\ref{sec:diagnostics}}
\end{figure}

We first address the question of whether a naked singularity forms after
the black hole and cosmological horizons collide
\cite{Gao:2008jv, Faraoni:2014nba,Faraoni:2012gz}.
The formation of a naked singularity would signal a breakdown of the
theory---either through the formation of a blowup in curvature,
or through necessitating new boundary conditions to be set at the 
singularity boundary \cite{1969NCimR...1..252P}.
Our simulations suggest no naked singularity is formed. 
More concretely, the outward null expansion
during this period is negative everywhere, so the entire
spacetime is essentially trapped, and no new boundary conditions
need to be specified.
In particular, we can continue to excise a central region corresponding to
the inside of the black hole.
Additionally, considering the event horizon shown in figure~\ref{fig:2m_AH}, 
we see that it remains finite at all times. 
Note that just like in the case studied earlier in section~\ref{sec:low_mass}, 
the event horizon is smaller than the apparent horizon before, and
during the bouncing phase, and turns around when it crosses the apparent horizon.

We next consider the behavior of the marginally (anti-)trapped tubes and
their signature, shown in figure~\ref{fig:2m_AH}.
Note that while the 
black hole MTT is spacelike and increasing in time before it merges 
with the cosmological horizon, 
when it reappears from the merger, 
its signature remains spacelike even though its area continues to decrease in time. 
Since the area of the black hole always increases in the outward radial direction, 
this implies that while the outward direction points into the future before the merger, 
it points into the past when it reappears. 
The black hole apparent horizon undergoes another signature change at the bounce
(indicated by the grey vertical solid line)
after which it behaves like the case studied above (i.e. the signature
of the horizon becomes timelike, and decreases as we evolve forwards in time).
Similarly, we find that the cosmological horizon follows the same trend 
as the case in section~\ref{sec:low_mass}, except for a brief period of time just 
before it merges with the black hole apparent horizon: here the horizon
signature becomes spacelike.
We see that the cosmological horizon and black hole apparent horizons have the
same signature when they annihilate and re-emerge. 

A natural question to ask is whether the collision of the
apparent horizons during the contraction phase is
an artefact of the particular matter model we use, 
or is a more general consequence of a contracting universe.
To explore this, we consider the same initial 
conditions as the ones used in figure~\ref{fig:2m_rho}, 
but now evolve only with the canonically normalized scalar field.
The results of this are plotted in figure~\ref{fig:chioff}. 
We find that during contraction, the apparent horizons, with and without
the presence of a ghost scalar field, behave in a similar fashion.
In both cases,
the black hole apparent horizon merges with the cosmological horizon at the same areal radius.
This is in line with our earlier observation that the black hole
horizon's size exceeds the cosmological horizon before the bouncing phase starts,
(i.e. before the ghost field has a significant impact on the evolution
of the system). 
The black hole and cosmological apparent horizon merge earlier by around
$t\sim 2 |H_{\rm min}|$ in the case of
contraction without the ghost scalar field. 
This is consistent with the notion that the accretion of the
ghost field should slow down the rate at which the black hole can grow in size,
which would delay the time of merger of the two horizon.

Finally, we note that the signature of the cosmological horizon becomes 
spacelike in this setup just before merging with the black hole horizon
for both cases.
During this phase of evolution, the cosmological horizon is a dynamical
horizon whose area \emph{decreases} with time.

\begin{figure}[h]
\centering
\includegraphics[width=.495\textwidth]{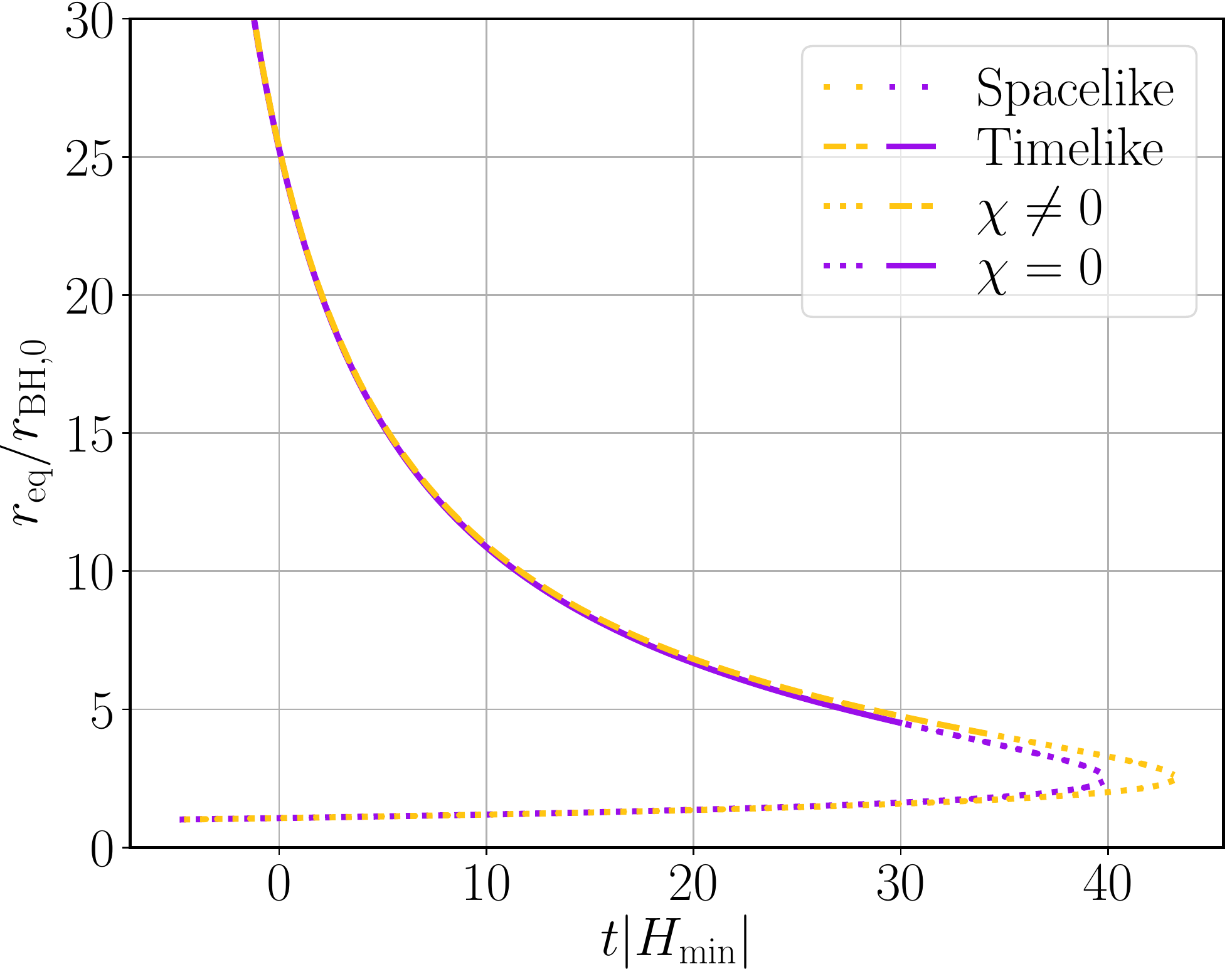}
\caption{\label{fig:chioff} 
   Cosmological and black hole apparent horizons from the contracting 
   phase of the same case shown in figure~\ref{fig:2m_AH} 
   (labelled $\chi \neq 0$) compared to a 
   similar case without a ghost field ($\chi = 0$).
}
\end{figure}

%-----------------------------------------------------------------------------
\subsection{Dependence on black hole size \label{sec:extrapolate}}
In this section, we explore in more detail how the properties of
the spacetime during the bounce change as a function 
of $R_{\rm{H},\rm{min}}/r_{\rm{BH},0}$.

As described in section~\ref{sec:low_mass}, 
for initial data where
$R_{\rm{H},\rm{min}} \approx 4.34 r_{\rm BH,0}$ 
the black hole apparent horizon persists through the whole bounce, 
and the spacetime evolution near the black hole 
qualitatively resembles the asymptotic cosmological evolution. 
As this behavior will hold to an even
greater degree for smaller black holes (relative to $R_{\rm{H},\rm{min}}$),
we are more interested in the opposite regime, considering larger black holes.
As mentioned in section~\ref{sec:intro}, for astrophysical black holes
we expect $R_{\rm{H},\rm{min}} \ll r_{\rm BH}$.
We find that, when $R_{\rm{H},\rm{min}} \lesssim 3.5 r_{\rm BH,0}$, 
(see section~\ref{sec:high_mass}),
the black hole apparent horizon collides with 
the cosmological horizon while the universe is still contracting.  
In this section, we therefore explore how this behaviour changes
as one increases the initial mass of the black hole.
We note that for 
numerical reasons\footnote{In particular,
we see a large growth in constraint violation near the outer boundary
when we increase the initial black hole mass to too large a size.
This is likely related
to the fact that we set our boundary conditions to be the
homogeneous cosmological solutions,
and that our runs lacked the resolution near the boundary
to resolved the correct falloff of the fields to their asymptotic values.},
we will restrict to evolutions where $R_{\rm{H},\rm{min}} > 0.86 r_{\rm BH,0}$.  
However, as we argue below, we already see some consistent
trends as $r_{\rm BH}$ is varied within this regime. 

In the left panel of figure~\ref{fig:trend}, 
we plot the radius of the black hole apparent
horizon normalized by its initial value as a function of time.
For evolutions where the black hole and cosmological
MTTs do not collide, we find that although the 
area of the apparent horizon always reaches it maximum
and minimum values at around the same harmonic time 
($t\sim 110 R_{\rm H,min}$ for the maximum value,
and $t\sim 440 R_{\rm H, min}$ for the minimum value), 
the value the maximum and minimum take does change as a function of initial
black hole area. Independently of the black hole's initial size,
the area of the apparent horizon is close to one around the bounce or in other words
around the time where the total
energy density of the background cosmology is zero.
In the low mass regime, the variation in the black hole's size increases
with increasing initial black hole area.

However as the initial size of the black hole increases,
the maximum change in the radius of the apparent horizon eventually
peaks at a value of $r_{\rm AH,max}/r_{\rm AH,0} \sim 2.6$.
In this case, the ratio of the minimum Hubble radius of background cosmology
to initial radius of the black hole corresponds to the threshold beyond which the
horizons merge.  Beyond this peak, although the horizons merge at successively 
earlier times (and always before the bouncing phase starts), 
with increasing initial black hole radius, the relative increase in the radius of the
apparent horizon when the horizons merge saturates at a value of 
$r_{\rm AH,max}/r_{\rm AH,0} \sim 2.5$. Within the range of masses we were able
to evolve, the apparent horizons always reappear, from which we conjecture that
the presence of black holes in bouncing cosmologies do not disrupt the bounce.
We were not able to evolve the space time to arbitrarily later proper time but 
based on all the simulations we have run, we conjecture that the black hole asymptotes
to close to its initial radius as $t \rightarrow \infty$ .

In the right panel of figure~\ref{fig:trend}, we plot the radius of the black hole event
horizon normalized by its initial value as a function of time.
We do not compute the evolution of the event horizon past the bounce for black holes with
initial black hole radius such that the minimum Hubble radius is smaller than 
$R_{\rm{H},\rm{min}} < 2.90 r_{\rm BH,0}$, 
as for those cases the event horizon cannot be located to the desired accuracy
(see appendix~\ref{sec:numerical_methodology} for more details on the 
computation of the event horizon). 
For the set of initial radii we do compute,
we find that the area of the event horizon 
reaches a maximum at successively earlier times 
with increasing initial black hole radius, 
always before the bouncing phase starts and always when the 
event horizon crosses the apparent horizon of the black hole. 
Beyond this point, the event horizon decreases in size,
until it crosses the apparent horizon again, after which it starts
increasing. This minimum happens at successively earlier times
with increasing initial black hole radius.
While the maximum size of the event horizon throughout the 
evolution increases with increasing initial black hole radius,
the minimum decreases. 

We next argue that the behavior of the event horizon in the region leading up 
to the bounce (where $H<0$), can be at least qualitatively captured 
by studying null rays in the background FLRW spacetime. 
The reasoning is as follows: 
It is reasonable to assume that in the regime
where $R_{\rm H}/r_{\rm BH,0} \gg 1$, 
the evolution of null rays near the black hole horizon
will not be greatly influenced by the background cosmological evolution.
Likewise, we assume that
in the regime where the black hole is ``large''
($R_{\rm H}/r_{\rm BH,0} \lesssim 1$),
the trajectories of null rays exterior to the black hole
are more influenced by the cosmological evolution\footnote{For example, 
we find that when the black hole and cosmological apparent horizons merge
(and thus there is no boundary between a trapped and untrapped region), 
the spacetime dynamics qualitatively resemble that of the 
background cosmological evolution.}, and 
in the background FLRW spacetime, during the contraction phase, outward radial
null rays have decreasing proper radius when they are inside the Hubble radius.
Following this line of thought, we integrate null rays backward in time
in the background FLRW spacetime given by eq.~\ref{eq:line_element_harmonic_cosmo}, 
starting from the latest time for which $H<0$ and $R_{\rm H}=r_{\rm BH,0}$. 
Figure~\ref{fig:null_rays} shows the trajectories of a few
such null rays for different ratios of $R_{\rm H,min}/r_{\rm BH,0}$.
We find that the proper radius of the null rays
increases (as we go backwards in time) until the ray crosses 
$R_{\rm H}$, after which it decreases.
This is consistent with the behavior of the event horizon in the right 
panel of figure~\ref{fig:trend} and suggests that, at least for this part
of the evolution, the size of the black hole is 
determined by the evolution of the background cosmology.
This simple calculation also shows that as $R_{\rm H,min}/r_{\rm BH,0}$
decreases, the maximum radius of the null ray increases.  This agrees with what
we see in our full numerical simulations.  Extrapolating this trend to
arbitrarily small $R_{\rm H,min}/r_{\rm BH,0}$ suggests that for arbitrarily
large black holes the peak of the event horizon will diverge.  However, we are
working with a cosmological solution that has undergone an infinite number of
e-folds of contraction to the past (see section \ref{sec:cosmo_id}).  If one
were to consider a bouncing model that had only had a finite period of contraction
(for example if we considered a cyclic cosmology \cite{Steinhardt:2001st}),
then the maximum of the event horizon would always be finite.

Evolving forward in time, into the region where the universe is expanding
($H>0$), we find that the event horizon continues to decrease until it crosses
the apparent horizon, at which point it begins to increase in size.  However,
this behavior can not be captured by integrating the null geodesics in the
background spacetime, which suggests that the influence of the black hole on
the geometry is more relevant when $H>0$, and for radii less than $r_{\rm
BH,0}$. Due to numerical issues, we are unable to evolve far enough in time to
determine if the minimum of the event horizon keeps decreasing and eventually
reaches a point where the event horizon ceases to exist as $R_{\rm
H,min}/r_{\rm BH,0} \rightarrow 0$.

Finally, we note that (as is shown in figure~\ref{fig:0p5m_AH}) one expects the
apparent and event horizons to converge to the same value at late times, but
for reasons mentioned earlier in this section, we are not able to evolve long
enough in time to show this happens for initial data with $R_{\rm{H},\rm{min}}
< 4.34 r_{\rm BH,0}$.

\begin{figure}[h]
\centering
\includegraphics[width=.495\textwidth]{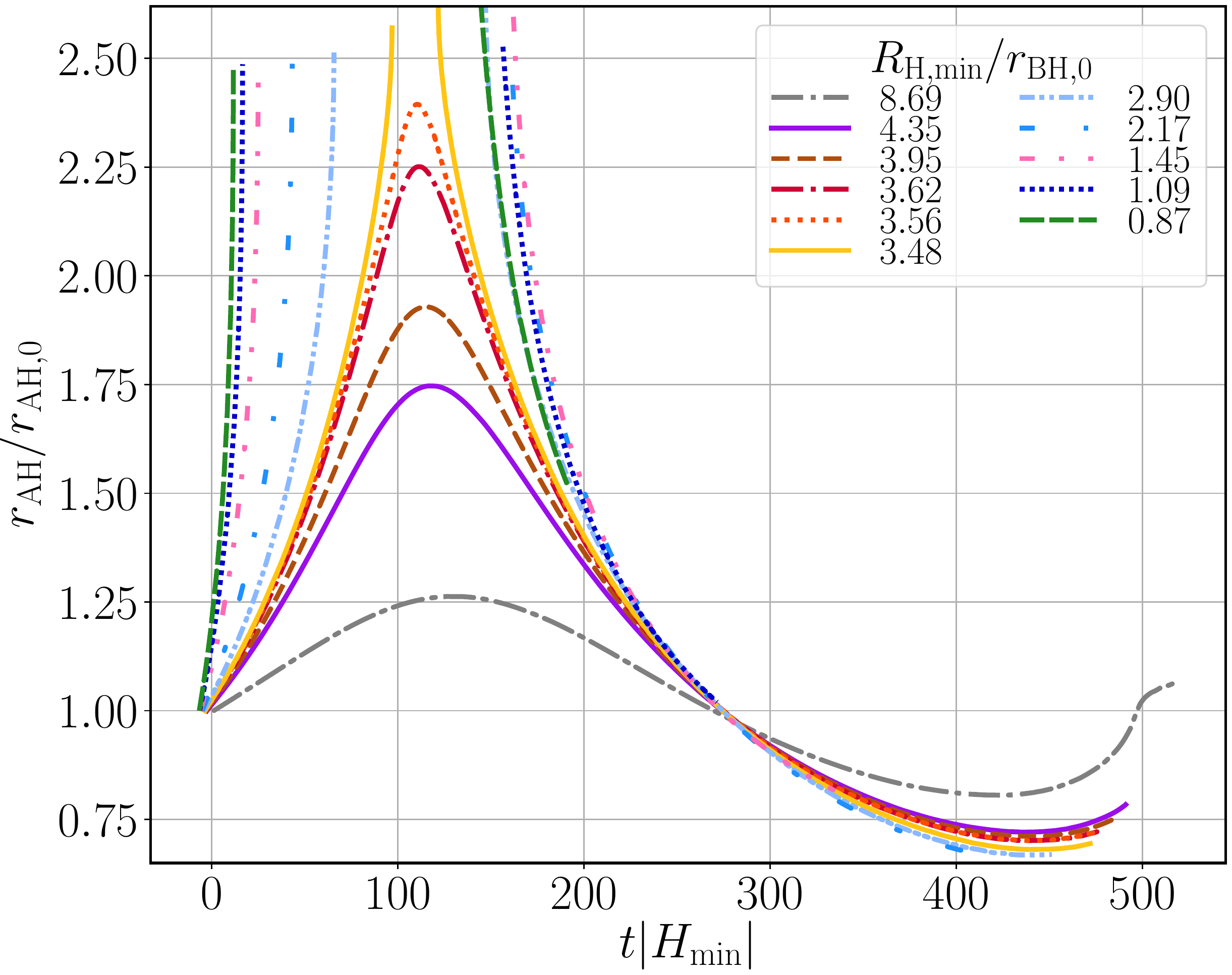}
\includegraphics[width=.495\textwidth]{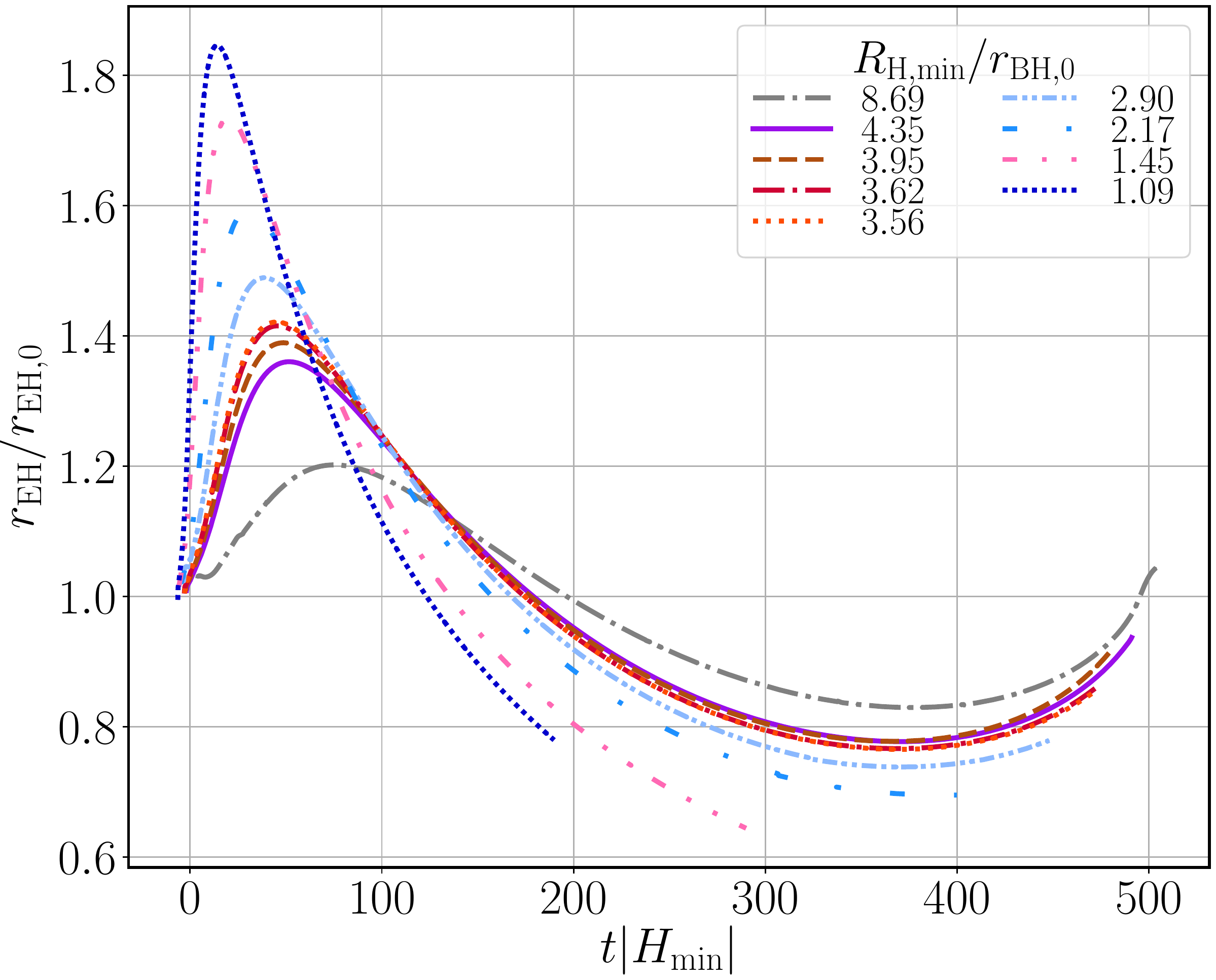}
\caption{\label{fig:trend} 
	Radius of the apparent (left) and event (right) horizon 
        of the black hole over time
	for different ratios of the minimum Hubble radius
        to initial black hole radius.
}
\end{figure}

\begin{figure}[h]
\centering
\includegraphics[width=.495\textwidth]{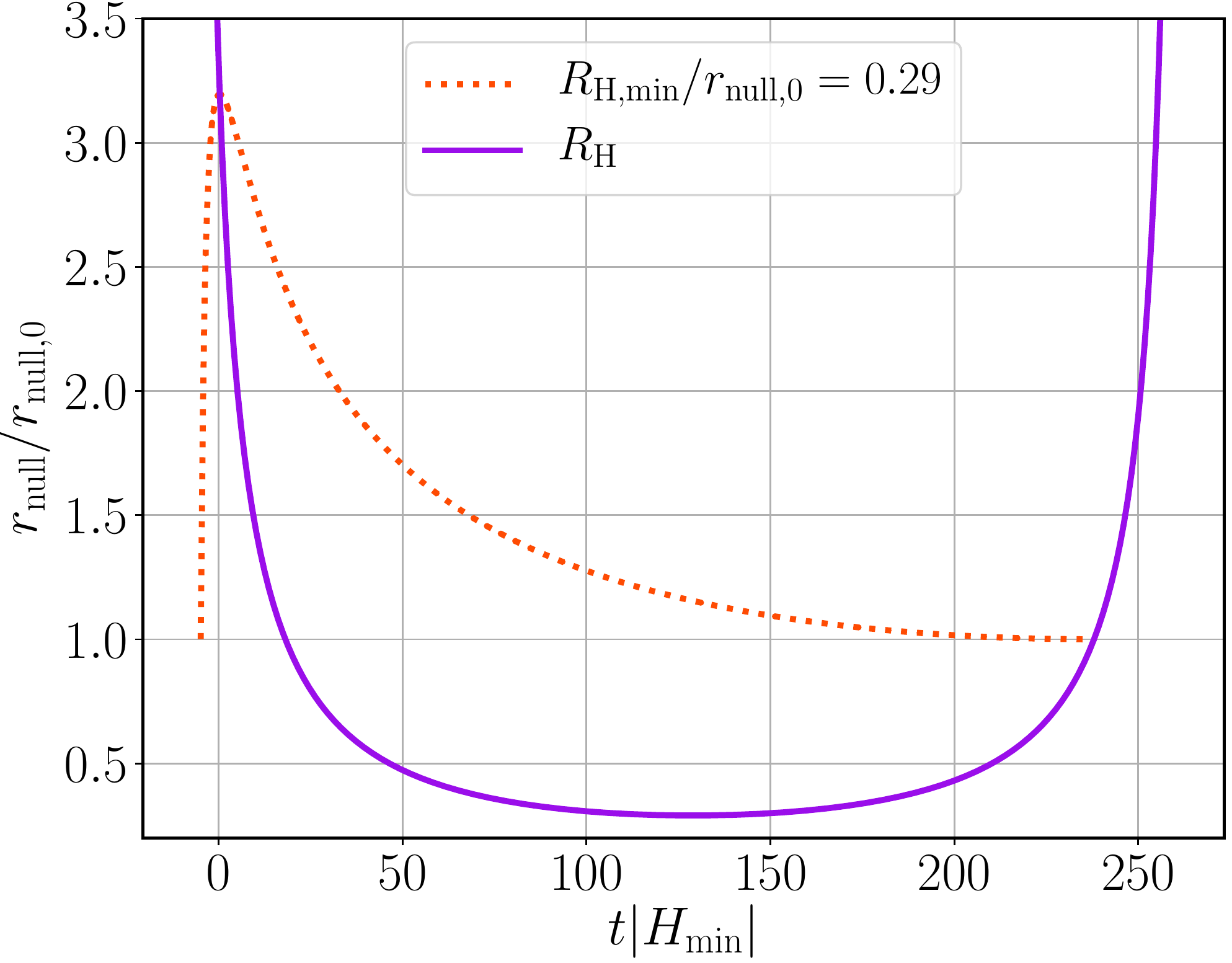}
\includegraphics[width=.495\textwidth]{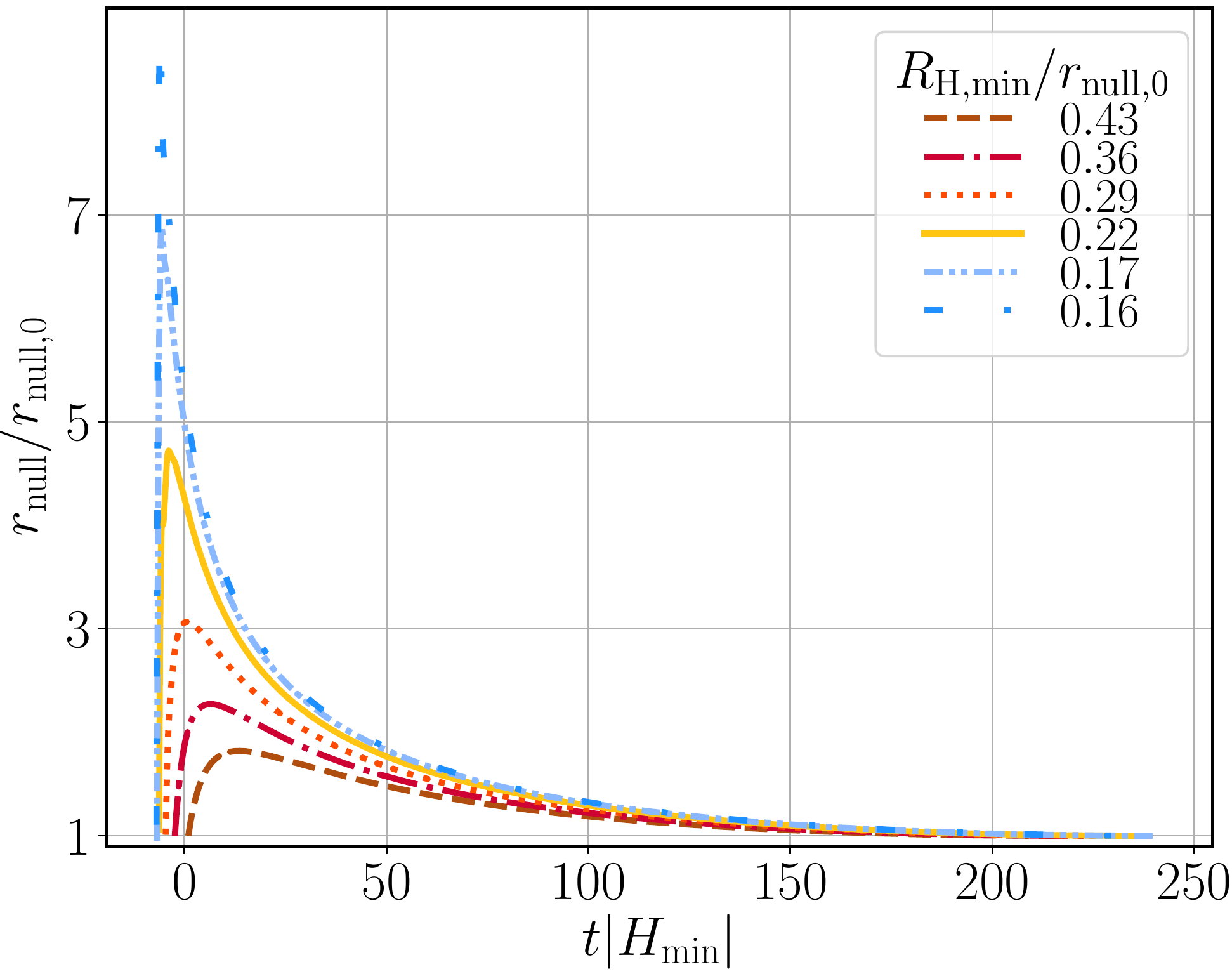}
\caption{\label{fig:null_rays} 
    Proper radius of outward, radial null rays in the background FLRW spacetime (described 
        in section~\ref{sec:equations_of_motion}).
        The left panel shows an example null ray that begins at a specified radius $r_{\rm null, 0}$, 
        increases until it crosses the Hubble radius (during contraction), and then decreases
        until it reaches the Hubble radius again at $r_{\rm null, 0}$.
        The right panel shows the same thing
        for different ratios of the minimum Hubble radius
        to $r_{\rm null, 0}$.
}
\end{figure}

%-----------------------------------------------------------------------------
\subsection{Spinning black holes\label{sec:spinning_bh}}
Up to this point, we have only considered 
non-spinning black holes (spherically symmetric spacetimes). 
However, our methods can be applied equally well to spinning black holes
spacetimes.
We have considered several such cases, finding the same 
qualitative behavior as for non-spinning black hole initial data.
We illustrate this with a representative example case:
initial data where the black hole is initially spinning with a 
dimensionless spin value of $a_0 = 0.5$. 
As we find little difference compared to the spacetimes
with non-spinning black holes, here we only present the results for a 
black hole with $R_{\rm H, min} = 2.17 r_{\rm BH,0}$,
and the same mass as in section~\ref{sec:high_mass}. 
Figure \ref{fig:spin} compares the circumferential radius
of the black hole along the equator for the spinning and non-spinning cases. 
We find that the addition of spin causes the horizons to 
merge at a slightly later time as compared to 
a comparable non-spinning case.
We do not plot the behavior of the asymptotic
background cosmology as it is the same regardless of 
whether the black hole is spinning or not. 
As was mentioned in section
~\ref{sec:diagnostics}, the angular momentum of the black hole
is constant since the scalar field
does not carry angular momentum.

\begin{figure}[h]
\centering
\includegraphics[width=.495\textwidth]{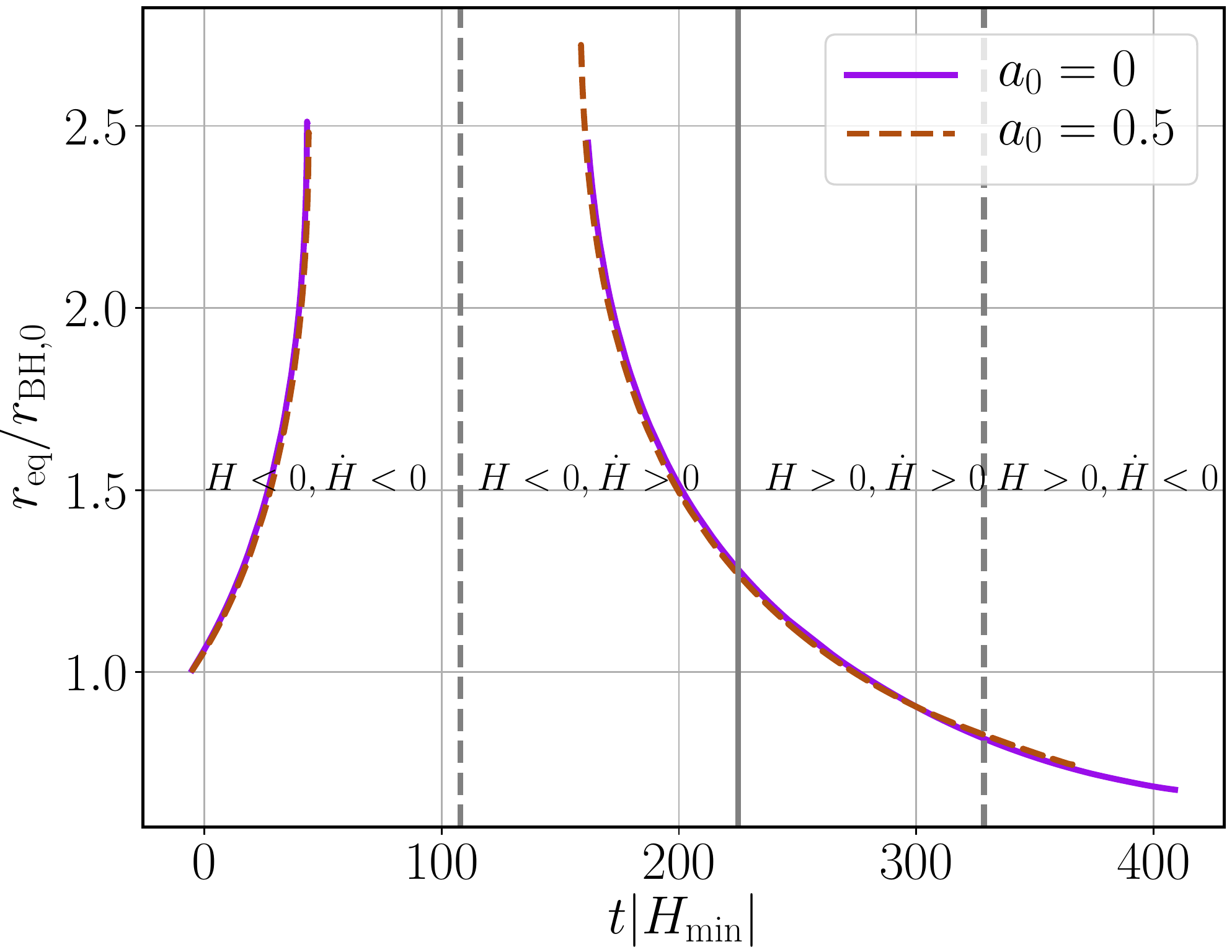}
\caption{\label{fig:spin} The circumferential radius of the black hole in 
   figure~\ref{fig:2m_AH} with zero spin ($a=0$) 
   compared to  dimensionless spin of $a=0.5$.
}
\end{figure}

%=============================================================================
\section{Discussion and conclusion\label{sec:discussion}}
We have considered the first numerical evolution of black holes through
a nonsingular bouncing cosmology.
As in~\cite{Allen:2004vz,Xue:2013bva}, we worked with a model that
has two scalar fields: a canonically normalized field with an exponential
potential and a ghost field.
We have additionally considered asymptotically cosmological
initial data that is tuned to allow for a matter-like
(effective equation of state $w=0$) contraction which is then followed
by a bounce that ends with cosmological expansion.
In~\cite{Xue:2013bva}, translational symmetries were assumed in two spatial
directions, which precludes the formation of black holes. 
By contrast, in this work we considered axisymmetric spacetimes
which allowed us to study the behavior of black holes through a bounce. 
While only a small fraction of Hubble patches are expected to have a black hole
during the late stages of ekpyrotic contraction 
\cite{Battefeld:2014uga,Chen:2016kjx,Quintin:2016qro}, our setup allows
us to examine the robustness of the ghost-field bounce, which in turn
serves as an effective classical model of NCC violation.

We found two qualitatively different kinds of spacetime evolution, which
depended on the ratio of the minimum Hubble radius of the background cosmology
to the initial radius of the black hole.  For black holes with initial radius
smaller than $\sim3.5$ times the minimum size of the Hubble radius of the
background cosmology, the black hole passes through the bounce freely and the
background cosmology remains largely unaffected (see
section~\eqref{sec:low_mass}).  Beyond this limit, we found that while regions
far away from the black hole still bounce freely, regions close to the black
hole evolve differently (see section~\eqref{sec:high_mass}).  In particular, we
found that during the contracting phase, the cosmological horizon and the black
hole apparent horizon merge and cease to exist for a brief period of time.
Some finite time later, before the bounce but after the background Hubble
radius reached its minimum size, the cosmological and black hole apparent
horizons separate.  Within the range of masses we considered, we found that the
black hole size (as measured by its horizon radius), varies significantly
during its evolution.  However, regardless of the initial mass of the black
hole, we found that the late time evolution consists of a black hole in an
expanding universe with a mass similar to its initial value.  Although we were
not able to evolve spacetimes where the Hubble radius shrinks to a much smaller
size compared to the radius of the black hole, we conjecture that the black
hole always survives through the bounce. This means that black holes created
(or already present) in the contraction
phase~\cite{Chen:2016kjx,Quintin:2016qro} can persist to have observational
consequences in the post-bounce era.

We found instances where the event and apparent horizons 
decrease as a result of our spacetime violating the NCC. 
Independently of the NCC being violated, we found
that in the regime where the black hole and
cosmological apparent horizon collide, the latter becomes
spacelike shortly before merging with the black hole. 
This is consistent with the observation that the
signature of the marginally (anti-)trapped tubes
changes such that any merging/reappearing 
pair of horizons always has the same signature.

Finally, we point out a few directions for future research.  One would be to
study the dynamics in a setup where the asymptotic cosmology is not prescribed.
For example, this could be accomplished by considering a toroidal/periodic
setup, and then considering a ``lattice'' of black holes
\cite{Bentivegna:2012ei,Clifton:2017hvg}. This setting would allow for the
study of the impact of black holes, as well as other perturbations on the
overall dynamics of the bounce.  While small perturbations have not been found
to appreciably change the dynamics of a nonlinear bounce when translational
symmetries are assumed \cite{Xue:2013bva}, it would be interesting to see if
perturbations could be more disruptive in the presence of a black hole, and in
a less-symmetric spacetime that does not preclude large-scale anisotropies.

Another direction would be to consider other models of cosmological bounces.
While we believe that the main conclusions we find here do
not depend strongly on the details of the bounce model, 
it would be interesting to determine
what differences would result from potentially more realistic models
of a bounce.
As we mention in the Introduction, the cosmological bounce scale
may be many orders of magnitude smaller than the initial size of a
primordial black hole. Due to the numerical instabilities
(as described in Sec.~\ref{sec:high_mass}), we were unable to carry out evolutions
in the regime $R_{\rm H,min}/R_{\rm BH,0}\ll 1$.
It would be interesting to see if our results still hold in this limit.

Another interesting question is the degree to which a ghost field, which can
reverse cosmic contraction, may similarly affect gravitational collapse and
singularity formation in a black hole interior.  NCC violating fields such as
ghost fields have been used to construct singularity free black hole-like
solutions, such as wormholes
\cite{doi:10.1119/1.15620,Visser:1995cc,Lobo:2007zb,Carvente:2019gkd}, so it is
not entirely implausible that there could be nontrivial dynamics near the
center of a black hole that accretes a ghost field.  In this study, we ignored
the dynamics deep inside the black hole, excising that region from our domain.
Exploring this would require coordinates better adapted to studying the
interior of black hole spacetimes, such as null coordinates
\cite{Brady:1995ni}.
%==============================================================================
\acknowledgments

We are grateful to Badri Krishnan, Frans Pretorius, Anna Ijjas and Paul Steinhardt for 
helpful discussions about bouncing cosmologies and dynamical horizons,
and to Alexander Vikman for pointing out several helpful references.
We are also grateful to the referee for their detailed and helpful review.
MC and WE acknowledge support from an NSERC Discovery grant.
JLR is supported by STFC Research Grant No. ST/V005669/1. 
This research was supported in part by 
Perimeter Institute for Theoretical Physics. 
Research at Perimeter Institute is supported in
part by the Government of Canada through the
Department of Innovation, Science and Economic Development
Canada and by the Province of Ontario through the
Ministry of Economic Development, Job Creation and
Trade. This research was enabled in part by support
provided by SciNet (www.scinethpc.ca) and Compute
Canada (www.computecanada.ca). Calculations were
performed on the Symmetry cluster at Perimeter Institute, the Niagara cluster at the University of 
Toronto, and the Narval cluster at Ecole de technologie sup\'erieure 
in Montreal.
This work also made use of the Cambridge Service for Data Driven
Discovery (CSD3), 
part of which is operated by the University of Cambridge Research
Computing on behalf of the STFC DiRAC HPC Facility (www.dirac.ac.uk)

%=============================================================================
\appendix
%=============================================================================
\section{
   Numerical methodology\label{sec:numerical_methodology}
}
We solve the equations of motion \eqref{eq:equations_of_motion}
using the generalized harmonic formulation as described
in \cite{Pretorius:2006tp}.
The numerical scheme we use follows that of~\cite{East:2011aa},
which we briefly summarize here.
We discretize the partial differential equations in space, using standard
fourth-order finite difference stencils, and in time, using fourth-order
Runge-Kutta integration.
We control high frequency numerical
noise using Kreiss-Oliger dissipation~\cite{1972Tell...24..199K}.
We use constraint damping to control the constraint
violating modes sourced by truncation error, with
damping parameter values similar to those used
in black hole evolutions using the generalized harmonic
formulation~\cite{Pretorius:2006tp}.
We fix the gauge freedom by working in harmonic coordinates, 
$\Box x^{\alpha}=0$.
During the expansion phase, 
we dynamically adjust the time step size in proportion to 
the decreasing global minimum of $1/\alpha$ where $\alpha$ is the lapse
(this would be $\alpha=1/a^3$ in a homogeneous FLRW universe,
see eq.~\eqref{eq:line_element_harmonic_cosmo}) 
in order to avoid violating the 
Courant-Friedrichs-Lewy condition~\cite{East:2015ggf,East:2017qmk}.

Following \cite{Pretorius:2004jg}, we dynamically track the outer apparent
horizon of the black hole, and excise an ellipsoid-shaped region interior
the horizon. We typically set the
ratio of the maximum ellipsoid axis to the maximum
black hole radial value to be $0.6$.

We compute the event horizon
by integrating null surfaces backwards in time 
\cite{Anninos:1994ay,Libson:1994dk,Thornburg:2006zb}
(we restrict this to spherically symmetric cases, where it
is sufficient to consider spherical null surfaces).
Since we are not able to evolve the spacetime to infinite proper time
(at which point the event and apparent horizon would coincide), 
we cannot precisely determine the final position of the event horizon. 
Instead, we use the apparent
horizon as the approximate location of the event horizon and choose a range of
initial guesses around this value. 
For two surfaces initially separated by $2.5
r_{\rm{BH},0}$, we find that their separation decreases to $0.1 r_{\rm{BH},0}$ within $\sim 4 \times 10^{-3}
|H_{\rm min}|^{-1}$ when evolving the null surfaces backwards in time,
 after which we consider the location of the event horizon to be
accurate to the desired accuracy.  Note that the separation rapidly decreases
when integrating backwards in time, a direct consequence of the divergence of the null geodesics 
going forward in time \cite{Libson:1994dk}.  

We additionally
make the use of compactified coordinates so that physical 
boundary conditions can be placed at spatial infinity \cite{Pretorius:2004jg}:
\begin{align}
   \label{eq:compactified_coordinate_relation}
   x^i
   =
   \tan \left(\frac{\pi \hat{x}^i}{2}\right)
   ,
\end{align}
so that $\hat{x}^i=1$ corresponds to $x^i=\infty$.
Unlike in \cite{Pretorius:2004jg} though, we work in an asymptotically
FLRW spacetime instead of an asymptotically flat spacetime,
similar to what is done in \cite{East:2016anr}.
That is, at our spatial boundary we set
\begin{align}
   g_{tt}
   =
   -
   \alpha(t)^2
   ,\qquad
   g_{ti}
   =
   0
   ,\qquad
   g_{ij}
   =
   a(t)^2\delta_{ij}
   ,
\end{align}
where the lapse $\alpha(t)=a(t)^3$;
and the scale factor $a(t)$ satisfies the Friedmann
equations, eq.~\eqref{eq:harm_ghost}.

We use Berger-Oliger~\cite{1984JCoPh..53..484B} style adaptive mesh refinement
(AMR) supported by the PAMR/AMRD library~\cite{Pretorius:2005ua,PAMR_online}.
Typically our simulations have 9--12 AMR levels (using a $2:1$ refinement ratio), 
with each nested box centered
on the initial black hole and between 128 and 256 points across the x-direction 
on the coarsest AMR level.
The interpolation in time for the AMR boundaries is only third-order accurate,
which can reduce the overall convergence to this order in some instances. 
As we restrict to axisymmetric spacetimes,
we use the modified Cartoon method
to reduce our computational domain to a
two-dimensional Cartesian half-plane~\cite{Pretorius:2004jg}.

We construct initial data describing a black hole of mass $M(t=0) = M_0$ 
in an initially contracting FLRW spacetime described in section~\ref{sec:cosmo_id}. 
We solve the constraint equations using the conformal thin 
sandwich formalism, as described in \cite{East:2012zn}. 
More precisely, we choose the initial time slice to 
have constant extrinsic curvature $K = - 3 H_0$ where 
$H_0 = \mathcal{H}_0 {a_0}^{-3}$ is given by~\eqref{eq:flrw_hamiltonian_constraint}, 
and the initial values for 
$\{ \phi,\phi',\chi,\chi',a,a'\}$ are fixed by \eqref{eq:cosmo_harm_id}
(a similar approach was employed in \cite{East:2015ggf,East:2016anr}). 
Without loss of generality, we choose the initial value of the ratio 
between the energy density of the $\phi$ and $\chi$ fields and $V_0$ to be 
such that 
during the contraction phase, the Hubble radius of the background cosmology
$R_H \equiv |H^{-1}|$ shrinks from an initial value of 
$R_H(t=0) = 75 r_{\rm BH,0}$ to $4.34 r_{\rm BH,0}$
(here $r_{\rm BH,0}$ is the initial black hole radius). 
We considered a range of initial black hole masses, keeping the initial ratio of Hubble 
to black hole radius to be 75, but changing the minimum Hubble to initial black hole 
radius ratio from $4.34 r_{\rm BH,0}$ all the way to $0.87 r_{\rm BH,0}$. 
We also study some black holes with an initial dimensionless spin of $a_0 = 0.5$.

Finally, we present a convergence test of our code and setup. 
In figure~\ref{fig:resln}, we present the time evolution of the apparent horizon
of the black hole 
and the norm of the constraint violations $C^{\alpha}\equiv \Box x^{\alpha}$ 
integrated over the coordinate radius $r \leq  265 M_0$, for a non-spinning black hole 
with initial mass such that $R_{\rm{H},\rm{min}} = 1.45 r_{\rm BH,0}$, 
for different numerical resolutions. 
For this case, the lowest resolution is 128 points across the x-direction on the 
coarsest AMR level with 10 levels of mesh refinement and a spatial resolution of 
$dx/M_0 \approx 0.004$ on the finest level. 
The medium and high resolutions correspond, respectively, to an increased resolution
of $3/2$ and $2\times$ that of the lowest resolution run. 
We find that the constraints converge to zero at roughly third order. 
This is because the convergence is dominated by the third order time interpolation 
on the AMR boundaries. The medium resolution in the convergence study is equivalent 
to the resolution we use for all the other cases studied here. 
We place the mesh refinement such that the radius of the black hole resides inside 
the finest AMR level initially. 
During the evolution, the mesh refinement is adjusted according to
truncation error estimates to maintain roughly the same level of error. 

\begin{figure}[h]
\centering
\includegraphics[width=.495\textwidth]{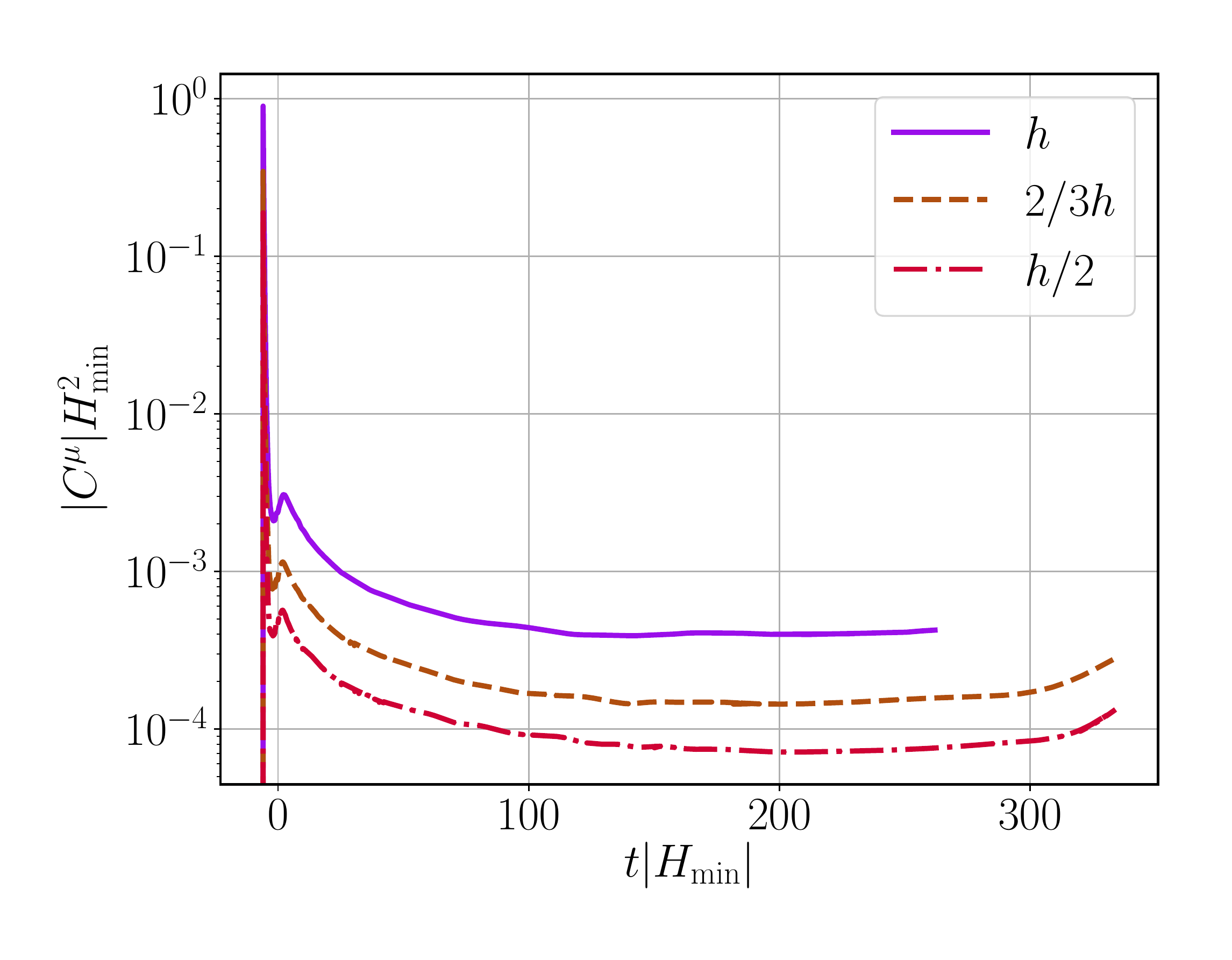}
\includegraphics[width=.495\textwidth]{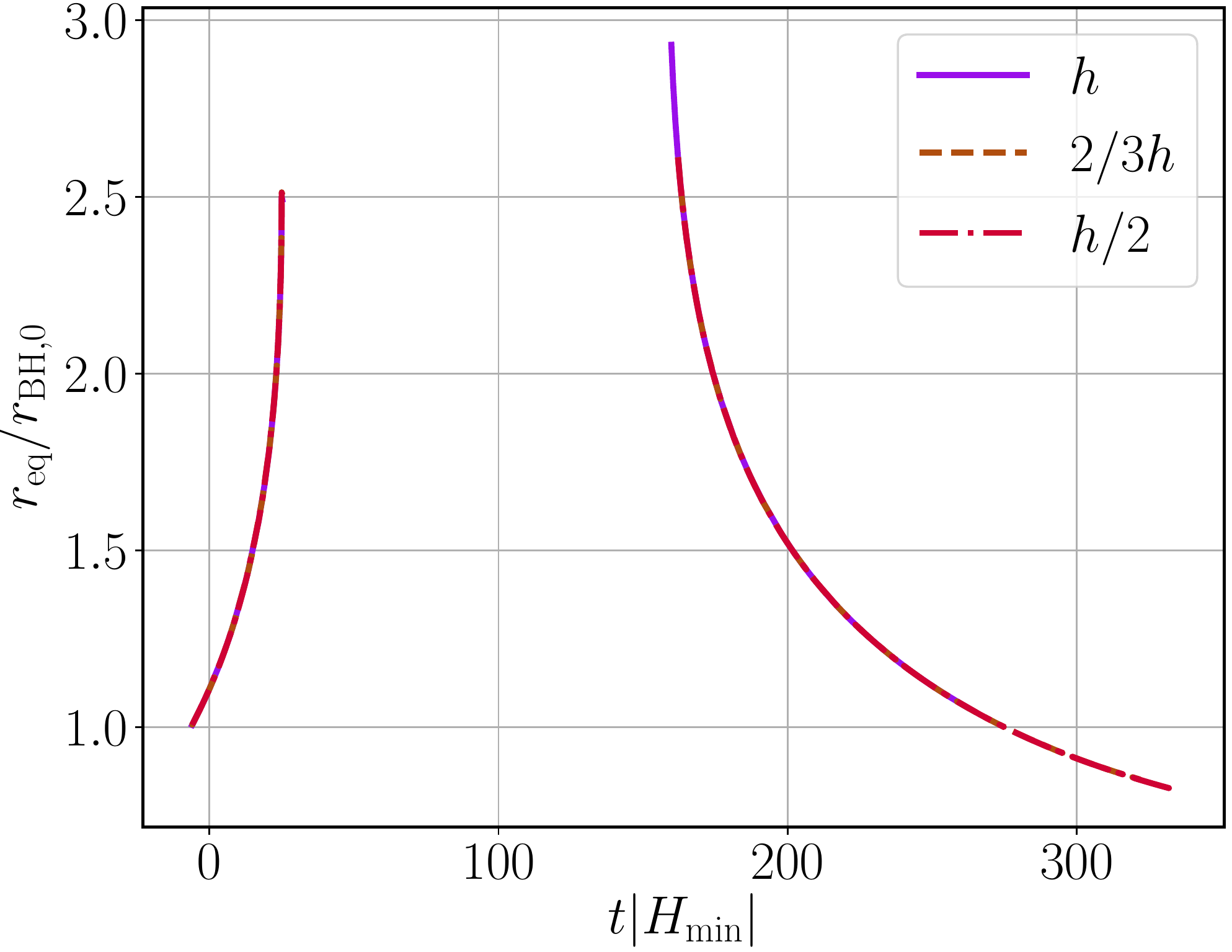}
	\caption{\label{fig:resln} The integrated norm (left) of the constraint violations 
	$C^{\alpha}\equiv \Box x^{\alpha}$ integrated over the coordinate radius 
	$r \leq  265 M_0$ and the apparent horizon (right), for a black hole with
	initial radius such that $R_{\rm{H},\rm{min}} = 1.45 r_{\rm BH,0}$, for different
	resolutions. The medium (high) resolution case has 
	$1.5 \times $ ($2 \times$) the resolution of the low resolution, and 
	the convergence is consistent with third order.
}
\end{figure}

\begin{figure}[h]
\centering
\includegraphics[width=.495\textwidth]{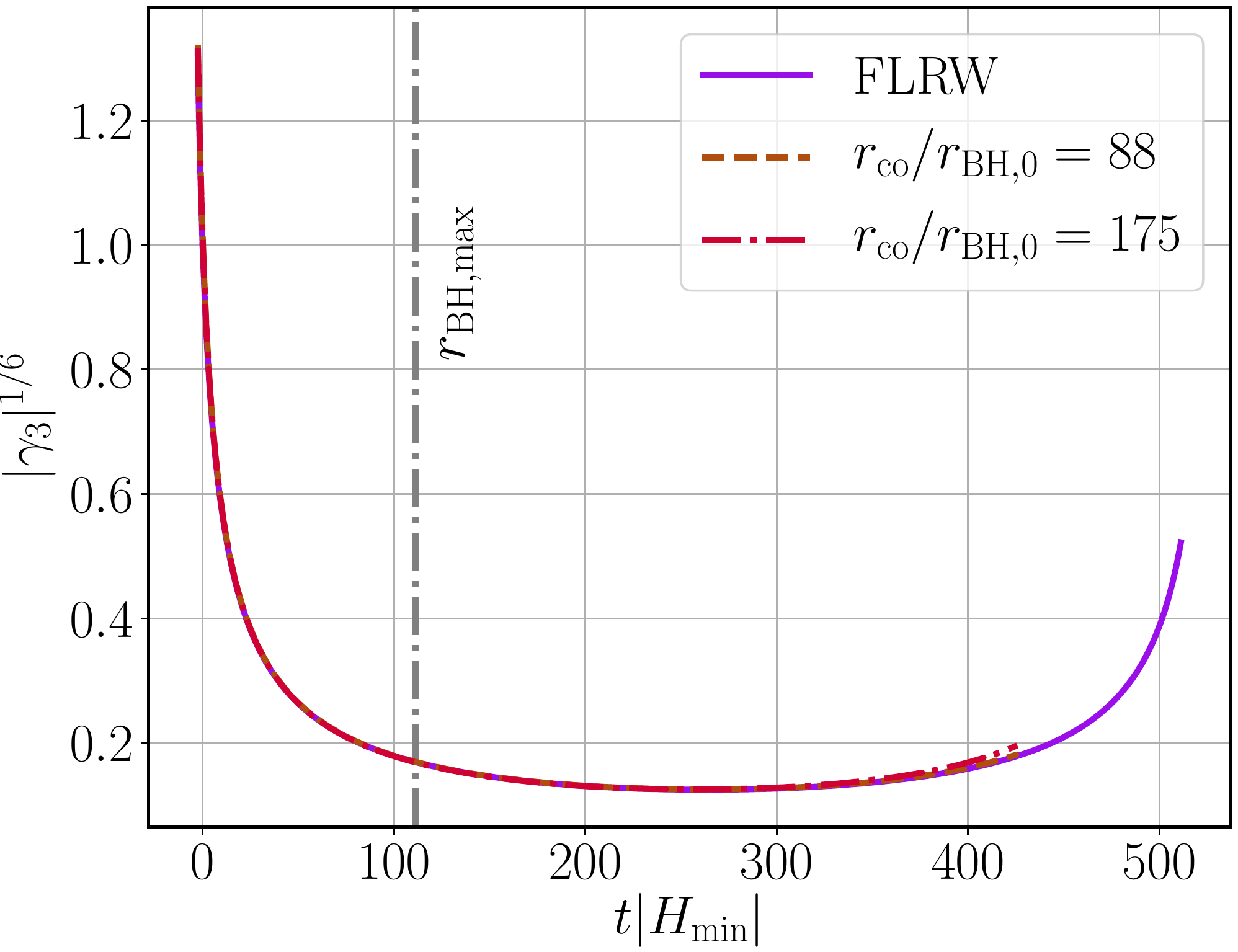}
\includegraphics[width=.495\textwidth]{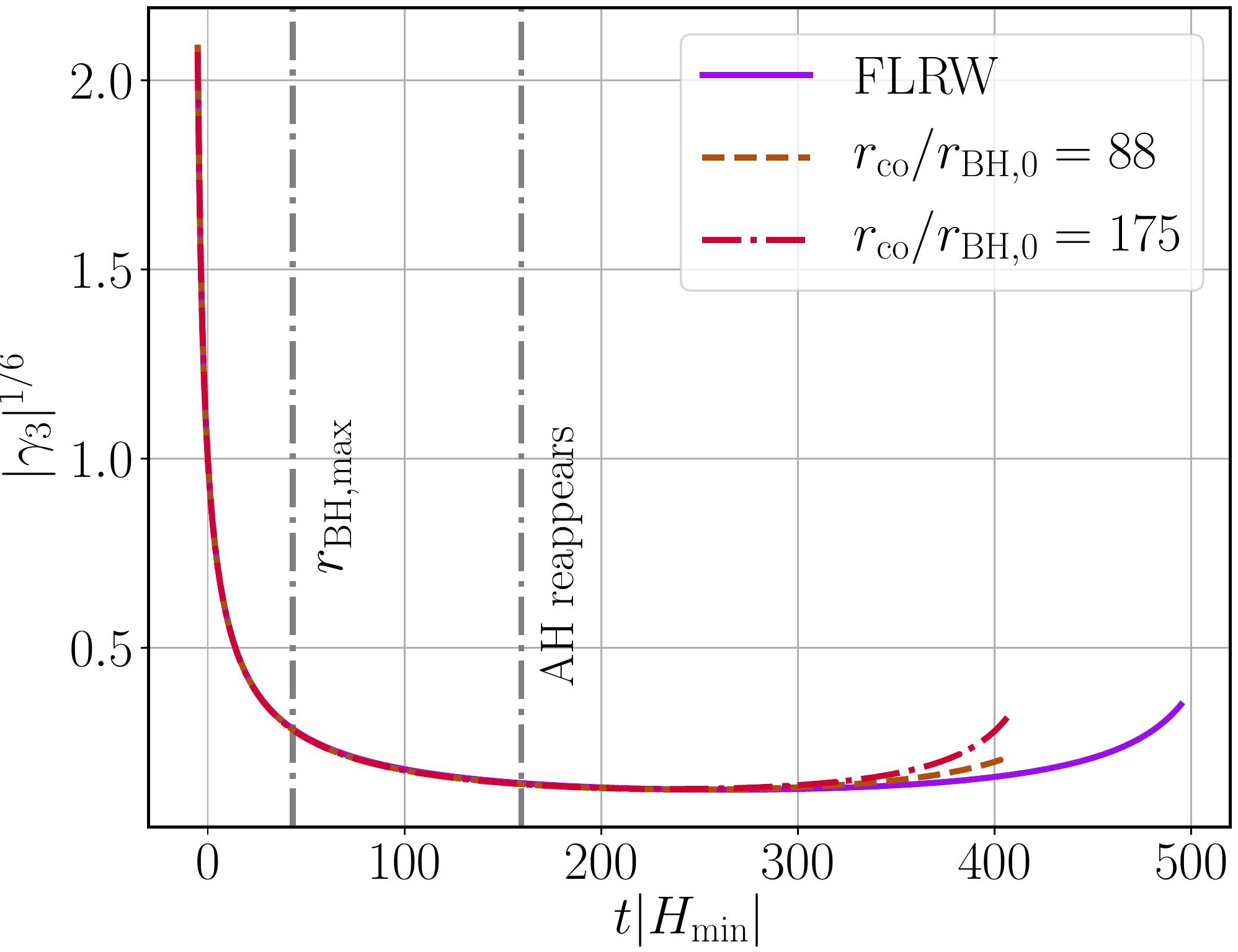}
\caption{\label{fig:aeff} 
The effective scale factor $|\gamma_{3}|^{1/6}$ computed from \eqref{eq:aeff} for a black hole 
	with initial mass such that the Hubble radius of the background cosmology
   $R_{\rm{H}} \equiv |H^{-1}|$ shrinks from an initial value of 
   $R_{\rm{H},0} = 75 r_{\rm BH,0}$ to $4.34 r_{\rm BH,0}$ (left)/$2.17 r_{\rm BH,0}$ (right). 
   The solid line shows the corresponding background 
   solution and the dashed and dash-dotted lines the values at 
   different coordinate radii. 
   The vertical grey line is the time at which the 
   black hole reaches its maximum mass as observed by the apparent horizon.
}
\end{figure}

\begin{figure}[h]
\centering
\includegraphics[width=.495\textwidth]{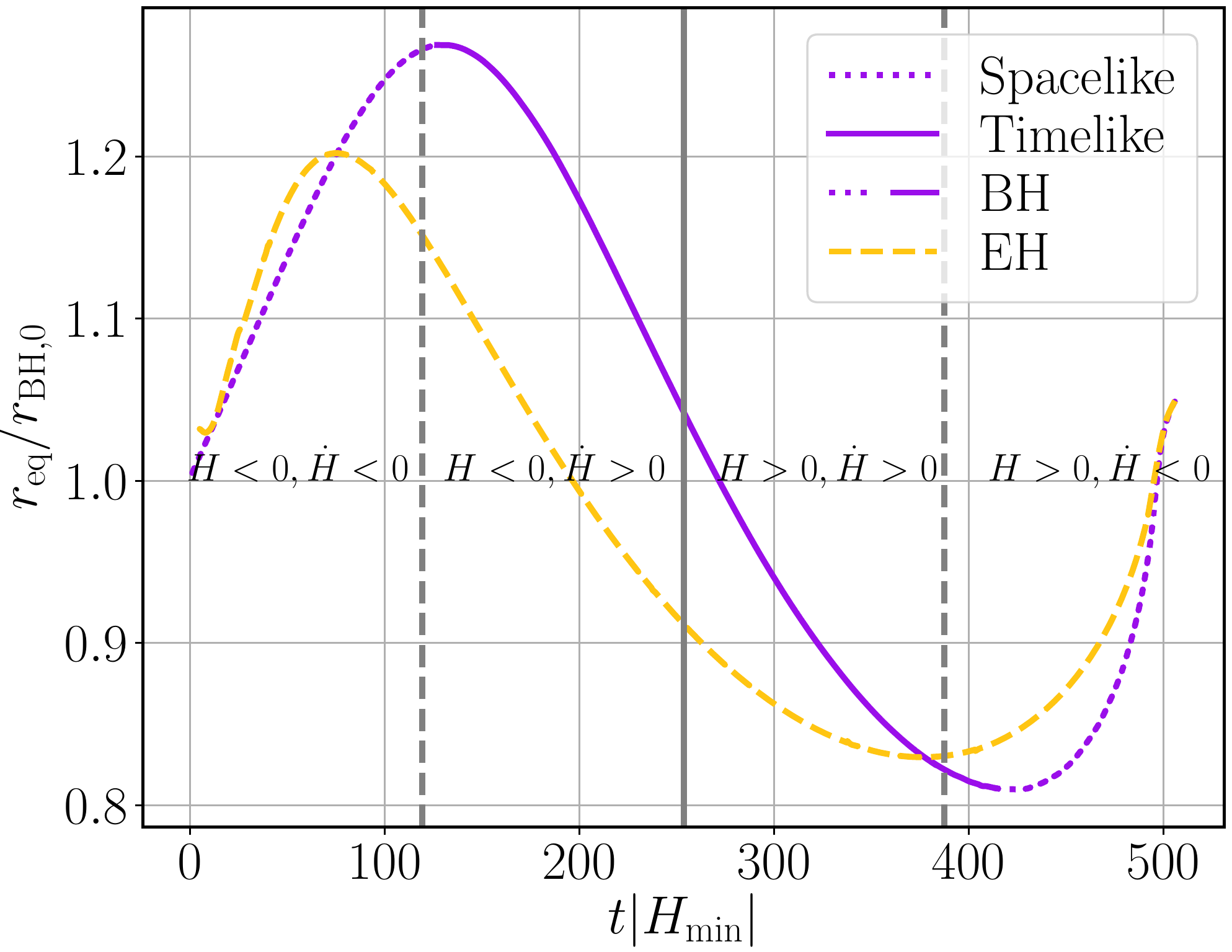}
\caption{\label{fig:0p5m_AH} 
   The apparent horizon of the black hole (purple) and 
   the corresponding event horizon (yellow dashed) for a black hole
	with initial mass such that the Hubble radius of the background cosmology
   $R_{\rm{H}} \equiv |H^{-1}|$ shrinks from an initial value of 
   $R_{\rm{H},0} = 75 r_{\rm BH,0}$ to $8.69 r_{\rm BH,0}$. 
   The vertical solid line indicates the bounce, while the region between 
   the dashed lines is the bouncing phase (where the NCC is violated).
   The line style of the apparent horizon reflects the signature of 
	the marginally trapped tube 
   or holographic screen (solid is timelike, dashed is spacelike). 
}
\end{figure}

%=============================================================================
\section{
   Various notions of black hole and cosmological horizons\label{sec:DH}
}
\subsection{General definitions and properties}

Nonsingular classically bouncing cosmologies require the violation of
the NCC 
\cite{Molina-Paris:1998xmn,Khoury:2001bz,Allen:2004vz,
Battefeld:2014uga,Brandenberger:2016vhg}.
The NCC plays a fundamental fundamental
role in the classical area law for black holes
\cite{hawking1973large,Bardeen:1973gs}.
Given this, we pay particular attention to the dynamics of the black hole
horizon in our simulations.
In addition to the event horizon 
(which can only be computed once the whole spacetime is known \cite{Thornburg:2006zb}), 
there are several other quasi-local
definitions of black hole horizons which we measure:
dynamical horizons 
\cite{Ashtekar:2004cn,Ashtekar:2003hk,Ashtekar:2002ag}, apparent
horizons \cite{Ashtekar:2004cn,Booth:2005ng,Hayward:1993mw,Ashtekar:2003hk,
Ashtekar:2002ag,Schnetter:2006yt,faraoni2015cosmological}, 
and holographic screens \cite{Bousso:1999cb,Bousso:2015mqa,Bousso:2015qqa} 
(also called marginally trapped tubes \cite{Ashtekar:2005ez}).

For completeness, we collect the definitions and some of the basic properties
of these horizons in this appendix.
Wherever applicable, we also discuss how these definitions can
be extended to define cosmological horizons.
We refer the reader to 
\cite{Ashtekar:2004cn,Booth:2005ng,Hayward:1993mw,Ashtekar:2003hk,Ashtekar:2002ag,
Schnetter:2006yt,faraoni2015cosmological,
Bousso:1999cb,Bousso:2015mqa,Bousso:2015qqa} for more
thorough reviews on this subject.
\begin{itemize}
\item[] \textbf{Trapped surfaces and apparent horizons} \\
      Let $\mathcal{S}$ be a smooth, closed, orientable spacelike two-dimensional 
      submanifold in a four-dimensional spacetime $(\mathcal{M},g_{ab})$. 
      We then define two linearly independent, future-directed, null vectors normal to 
      $\mathcal{S}$, normalized\footnote{This convention varies across the literature. 
      We follow \cite{Schnetter:2006yt}, which is different from \cite{Ashtekar:2002ag}.} 
      such that
      \begin{equation}
      g_{\alpha\beta}l^{\alpha}n^{\beta} = -1
      \end{equation}
      where by convention $l^{\alpha}$ and $n^{\beta}$ 
      are respectively the outgoing and ingoing\footnote{In asymptotically flat 
      or AdS spacetimes, the notions of 
      \emph{outward} and \emph{inward} are the intuitive ones 
      but in cosmological spacetimes--where an independent notion of ``outward'' such 
      as conformal infinity do not exist--this is no longer true. 
      In the context of our numerical simulations, 
      we naturally define the outgoing (ingoing) direction as the direction 
      from the origin (spatial infinity) to spatial infinity (the origin).}  
      null normals.
      The two-metric induced on $\mathcal{S}$ is 
      \begin{equation}
         \tilde{q}_{\alpha\beta} 
         = 
         g_{\alpha\beta} +l_{\alpha} n_{\beta} +n_{\alpha} l_{\beta}
         ,
      \end{equation}
      and the null expansions are defined as
      \begin{equation}\label{eq:expansion}
         \Theta_{(l)} 
         \equiv 
         \tilde{q}^{\alpha\beta} \nabla_{\alpha} l_{\beta}
         ,
         \qquad 
         \Theta_{(n)} 
         \equiv
         \tilde{q}^{\alpha\beta} \nabla_{\alpha} n_{\beta}
         .
      \end{equation}
      The closed two-surface $\mathcal{S}$ is a 
      \emph{trapped surface} 
      if the outgoing and ingoing expansions are strictly
      negative i.e. if $\{\Theta_{(l)}<0, \Theta_{(n)} < 0 \}$ and a \emph{marginal
      trapped surface} if the outgoing null expansion vanishes i.e if $\{\Theta_{(l)}=0,
      \Theta_{(n)} < 0 \}$. Conversely, the closed two-surface $\mathcal{S}$ is am
      \emph{anti-trapped surface} if the outgoing and ingoing expansions are strictly
      positive i.e. if $\{\Theta_{(l)}>0, \Theta_{(n)} > 0 \}$ and a 
      \emph{marginal anti-trapped surface} if the ingoing null expansion vanishes i.e 
      if $\{\Theta_{(l)}>0, \Theta_{(n)} = 0 \}$.

      A marginal trapped surface ($\Theta_{(l)} = 0$) is \emph{future} if $\Theta_{(n)} < 0$,
      \emph{past} if $\Theta_{(n)} > 0$, \emph{outer} if $\mathcal{L}_{(n)} \Theta_{(l)} < 0$
      and \emph{inner} if $\mathcal{L}_{(n)} \Theta_{(l)} > 0$. 
      Conversely, a marginal anti-trapped surface ($\Theta_{(n)} = 0$) 
      is \emph{future} if $\Theta_{(l)} < 0$, \emph{past}
      if $\Theta_{(l)} > 0$, \emph{outer} if $\mathcal{L}_{(l)} \Theta_{(n)} < 0$ and
      \emph{inner} if $\mathcal{L}_{(l)} \Theta_{(n)} > 0$.\\ 
      
      A black hole \emph{apparent horizon} is a \emph{future marginally outer trapped surface}. 
      Within the context of cosmology, the \emph{cosmological apparent horizon} 
      of an expanding FLRW
      spacetime is a \emph{past marginally inner anti-trapped surface}.
      For a contracting FLRW spacetime, the \emph{cosmological apparent horizon} is a 
      \emph{future marginally inner trapped surface}. 

\item[] \textbf{Dynamical horizons and holographic screens} \\
      We now have all the ingredients to introduce the concept of a \emph{Marginally Trapped
      Tube} (MTT): A MTT is a smooth, three-dimensional submanifold that is foliated by MTSs.
      If a MTT is everywhere spacelike, it is referred to as a \emph{dynamical horizon}
      \cite{Ashtekar:2002ag,Ashtekar:2003hk}.
      If it is everywhere timelike, it is called a \emph{timelike
      membrane} (TLM)\footnote{More recently this surface has also been called
      a \emph{timelike dynamical horizon}; see appendix B of \cite{Ashtekar:2003hk}.}.
      Finally, if it is everywhere null then we have an isolated horizon.
      
      We next outline the various ingredients that 
      go into the area law of dynamical horizons (the quasi-local
      horizon that appears the most frequently in our numerical solutions). 
      It is straightforward to derive an area law for purely spacelike
      or purely timelike dynamical horizons.
      Consider first the spacelike case.
      Let $H$ be a dynamical horizon and $\mathcal{S}$ 
      be a member of the foliation of \emph{future marginally trapped surfaces}. 
      Since $H$ is spacelike, we can define a future-directed unit timelike 
      vector normal to $H$, $\hat{\tau}^a$ and a unit outward pointing 
      spacelike vector tangent to $H$ and normal
      to the cross-sections of $H$, $\hat{r}^{\alpha}$. 
      A suitable set of null normals is then
      \begin{equation}\label{eq:null_vectors}
         l^{\alpha} 
         = 
         \frac{1}{\sqrt{2}}(\hat{\tau}^{\alpha} +\hat{r}^{\alpha})
         ,
         \ \ \ 
         n^{\alpha} 
         = 
         \frac{1}{\sqrt{2}}(\hat{\tau}^{\alpha} -\hat{r}^{\alpha})
         .
      \end{equation}
      Then since (by the definition of a dynamical horizon)
      $\Theta_{(l)} =0$ and $\Theta_{(n)} < 0$,
      it follows that the extrinsic curvature scalar of $\mathcal{S}$ is
      \begin{equation}
         \tilde{K} 
         = 
         {\tilde{q}}^{\alpha\beta} D_{\alpha} \hat{r}_{\beta} 
         = 
         -
         \frac{\Theta_{(n)}}{\sqrt{2}} 
         > 
         0
         ,
      \end{equation}
      where $D_{\alpha}$ is the covariant derivative operator on $H$. 
      This shows that the area of the cross-sections of a spacelike dynamical 
      horizon increases along $\hat{r}^{\alpha}$. 
      We emphasize that this does not necessarily imply that the area 
      increases in time. 
      In spherical symmetry, we explicitly show below 
      (section~\ref{sec:dynamical_horizons_spherical_symmetry}) 
      that the outward vector points in the future when the area increases 
      in time and in the past when the area decreases in time. 
      For a timelike dynamical horizon, the roles of $\hat{\tau}^{\alpha}$ 
      and $\hat{r}^{\alpha}$ are interchanged. 
      In this case, $\hat{r}^{\alpha}$ is no longer tangential to $H$, 
      and is instead the unit spacelike vector normal to $H$. 
      Additionally, $\hat{\tau}^{\alpha}$ is instead the unit 
      timelike vector tangent to $H$ and orthogonal to the cross-sections of $H$. 
      The area law then becomes
      \begin{equation}
         \tilde{K} 
         = 
         \tilde{q}^{\alpha\beta} D_{\alpha} {\hat{\tau}}_{\beta} 
         = 
         +\frac{\Theta_{(n)}}{\sqrt{2}} 
         < 
         0
         ,
      \end{equation}
      i.e. the area of a timelike dynamical horizons 
      decreases along ${\hat{\tau}}^{\alpha}$.
      Note this law does not rely on any
      energy conditions, such as the NCC.

      Finally, we note that the area law defined in 
      \cite{Ashtekar:2002ag,Ashtekar:2003hk} only
      applies to dynamical horizons and timelike membranes.
      The definition does not include marginally
      anti-trapped tubes, which are often present in cosmological settings,
      or marginally trapped tubes which may not have a definite 
      signature at a given time. 
      To remedy this, Bousso and Engelhardt \cite{Bousso:2015mqa,Bousso:2015qqa}
      formulated and proved a new area theorem applicable to an entire hypersurface $H$ of
      indefinite signature. 
      The area theorem is based on a few technical assumptions 
      but should be applicable to most hypersurfaces
      foliated by marginally trapped or anti-trapped surfaces $\mathcal{S}$, 
      called \emph{leaves}. In this context, marginally 
      (anti-)trapped tubes are referred to as \emph{future (or past) holographic
      screens}. 
      More precisely, the Bousso-Engelhardt area theorem is:
      the area of the leaves of any future (past) holographic screen, $H$, increases
      monotonically along $H$. The direction of increase along a future (past) holographic
      screen is the past (future) on timelike portions of $H$ or exterior on spacelike
      portions of $H$. Thus $H$ only evolves into the past (future)
      and/or exterior of each leaf. 

\end{itemize}

%=============================================================================
\subsection{Dynamical horizons and timelike membranes in spherical symmetry
\label{sec:dynamical_horizons_spherical_symmetry}}

As most of our simulations are performed in an essentially spherically
symmetric spacetime, 
here we consider the properties
of dynamical horizons for these spacetimes in more detail.
The main purpose of this section is to illustrate 
how the area law for dynamical horizons 
\cite{Ashtekar:2002ag,Ashtekar:2003hk}
reduces to an essentially
tautological statement about the dynamics of the horizon area.

We use $r$ to denote the areal radius, and we will work with a gauge 
such that $r$ is also a coordinate of the spacetime,
that is we will consider a metric of the form
\begin{align}
   ds^2
   =
   \alpha_{ab}dx^adx^b
   +
   r^2\left(d\vartheta^2+\sin^2\theta d\varphi^2\right)
   ,
\end{align}
where $\alpha_{ab}$ is a two-dimensional metric that is function of $(t,r)$
(here $t$ is the timelike coordinate).
We recall that in spherical symmetry 
the expansion for a null vector $v^{\mu}$ is
\cite{baumgarte_shapiro_2010}
\begin{align}
   \Theta_{(v)}
   =
   \frac{1}{4\pi r^2}v^{\alpha}\partial_{\alpha}\left(4\pi r^2\right)
   =
   \frac{2}{r}v^r
   .
\end{align}
The last expression follows from our imposing a gauge such
that the areal radius is also a coordinate of our spacetime.

We consider the level sets of a function
\begin{align}
   F(t,r)
   \equiv
   \mathfrak{r}(t)-r
   .
\end{align}

\begin{itemize}
   
\item[] \textbf{Case 1: the level sets of $F$ are spacelike.}
   
   We define a unit timelike vector orthogonal to the level set of $F$:
   \begin{align}
      \hat{\tau}_{\alpha}
      \equiv&
      \frac{1}{\sqrt{-\nabla^{\beta}F\nabla_{\beta}F}}\nabla_{\alpha}F
      =
      \frac{1}{\mathcal{N}_{\hat{\tau}}}\left(\dot{\mathfrak{r}},-1,0,0\right)
      ,
   \end{align}
   where we have defined 
   $\mathcal{N}_{\hat{\tau}}\equiv\sqrt{-\nabla_{\alpha}F\nabla^{\alpha}F}$.
   We next find the unit spacelike vector orthogonal to $\hat{\tau}_{\alpha}$,
   $\hat{r}_{\alpha} \hat{r}^{\alpha}
      =
      1
      ,
      \hat{r}_{\alpha} \hat{\tau}^{\alpha}
      =
      0
      .$
   We write $\hat{r}^{\alpha}$ as
   \begin{align}
      \hat{r}^{\alpha}
      =
      \frac{1}{\mathcal{N}_{\hat{r}}}\left(\frac{1}{\dot{\mathfrak{r}}},1,0,0\right)
      ,
   \end{align}
   where $\mathcal{N}_{\hat{r}}$ is the normalization. 
   Defining the null vectors according to \eqref{eq:null_vectors},
   a surface $\mathfrak{r}(t)$ is trapped if 
   $\Theta_{(l)}=0,\Theta_{(n)}<0$, and it is anti-trapped if 
   $\Theta_{(l)}>0,\Theta_{(n)}=0$.
   The area law for dynamical horizons states
   that the area of the
   dynamical horizon must increase in the direction of $\hat{r}^{\alpha}$
   as we evolve along $\hat{r}^{\alpha}$ \cite{Ashtekar:2002ag,Ashtekar:2003hk}.
   From the form of $\hat{r}^{\alpha}$, we see
   that this reduces to: if $\dot{\mathfrak{r}}>0$,
   then the dynamical horizon area increases 
   in the direction of increasing time,
   and if $\dot{\mathfrak{r}}<0$, then the 
   dynamical horizon areas increases in the direction of
   decreasing time.

\item[] \textbf{Case 2: the level sets of $F$ are timelike.}

   Analogous to the case when the level set is spacelike,
   we define a unit spacelike vector orthogonal to the level set of $F$:
   \begin{align}
      \hat{r}_{\alpha}
      \equiv&
      \frac{1}{\sqrt{-\nabla^{\beta}F\nabla_{\beta}F}}\nabla_{\alpha}F
      =
      \frac{1}{\mathcal{N}_{\hat{r}}}\left(\dot{\mathfrak{r}},-1,0,0\right)
      ,
   \end{align}
   and a unit timelike vector orthogonal to $\hat{r}_{\alpha}$,
   \begin{align}
      \hat{\tau}^{\alpha}
      =
      \frac{1}{\mathcal{N}_{\hat{\tau}}}\left(\frac{1}{\dot{\mathfrak{r}}},1,0,0\right)
      ,
   \end{align}
   where $\mathcal{N}_{\hat{t}}$ is the normalization. 
   Again, a surface $\mathfrak{r}(t)$ is trapped if 
   $\Theta_{(l)}=0,\Theta_{(n)}<0$, and it anti-trapped if 
   $\Theta_{(l)}>0,\Theta_{(n)}=0$.
   The area law for timelike membranes states
   that the area of the timelike membrane must decrease 
   in the direction of $\hat{\tau}^{\alpha}$
   as we evolve along $\hat{\tau}^{\alpha}$ \cite{Ashtekar:2002ag,Ashtekar:2003hk}.
   From the form of $\hat{\tau}^{\alpha}$, we see that this statement
   then reduces to: if $\dot{\mathfrak{r}}>0$,
   then the membrane area increases in the direction of increasing time,
   and if $\dot{\mathfrak{r}}<0$, then the membrane area increases 
   in the direction of decreasing time.

\end{itemize}
%=============================================================================
\section{
   The McVittie spacetime\label{sec:mcvittie}
}
   Here we briefly review the McVittie
spacetime \cite{original_mcvittie_paper}
(see also \cite{Nolan:1998xs,Nolan:1999kk,Nolan:1999wf,Li:2006zh,
Faraoni:2007es,Kaloper:2010ec,Faraoni:2012gz}), 
which is an analytic
solution to the Einstein equations that describes a spherically symmetric
black hole embedded in an asymptotically cosmological 
spacetime provided the cosmology asymptotes (in time: $t \to \infty$) 
to a de-Sitter
cosmology--for more discussion on this point, 
see \cite{Kaloper:2010ec}\footnote{We note that while other spacetimes 
that describe black
holes embedded within an asymptotically FLRW universe have been
proposed \cite{Schucking1954-zf,Sussman1985-zo,Gibbons:2009dr}, 
here we only focus on the McVittie
spacetime as it remains the most widely studied 
exact spacetime of this form.}.
The two most salient properties of the McVittie spacetime are that 
the spacetime is spherically symmetric and satisfies the 
\emph{no-accretion condition}, ${G^r}_t = 0$, which in turn implies that the 
stress-energy component ${T^r}_t = 0$. Thus, there is no radial
flow of cosmic fluid in the McVittie solution
(this assumption can be dropped for some generalizations
of the McVittie spacetime \cite{Faraoni:2007es}).
We relax all of these assumptions in our numerical simulations, in addition
to working in a set of coordinates that allows us to extend our
spacetime past the black hole horizon, which to our knowledge
has not yet been accomplished for the McVittie spacetime or its generalizations.
While our numerical solutions differ in many of their properties
from the McVittie spacetime, the McVittie spacetime serves as a useful
analytic example to understand some of the properties
of dynamical, apparent, and event horizons in spacetimes
that have a black hole and an asymptotic cosmological expansion
(see section~\ref{sec:DH}).

We consider only spatially flat McVittie solutions.
The spacetime metric in isotropic coordinates is
\begin{align}
   \label{eq:mcvittie_metric0}
   g_{\alpha\beta}dx^{\alpha}dx^{\beta}
   =&
   -
   \frac{\left(1-\frac{m(t)}{2\bar{r}}\right)^2}
   {\left(1+\frac{m(t)}{2\bar{r}}\right)^2}dt^2
   +
   a^2(t) \left(1+\frac{m(t)}{2\bar{r}}\right)^4
   \left( d\bar{r}^2 +\bar{r}^2 d\Omega^2 \right)
   .
\end{align}
where the McVittie no-accretion condition requires that the mass function $m(t)$
satisfies
\begin{align}\label{eq:no_accretion}
\frac{\dot{m}}{m} = -\frac{\dot{a}}{a}
\end{align}
or 
\begin{align}\label{eq:no_accretion_mass}
m(t) = \frac{m_H}{a(t)}
\end{align}
where $a(t)$ is the scale factor of the FLRW background, an overdot denotes
differentiation with respect to comoving time, and $m_H\geq0$
is an integration constant.
The Misner-Sharp \cite{misner_sharp_mass_paper}
(or Hawking-Hayward \cite{Hawking:1968qt,Hayward:1994bu}) quasi-local mass
$M_{MS}$---which is a coordinate invariant notion of 
energy for spherically symmetric spacetimes---is defined to be
\begin{align}
   \label{eq:def_ms_mass}
   1
   -
   \frac{2M_{MS}}{r_A}
   \equiv
   \left(\nabla r_A\right)^2
   ,
\end{align}
where $r_A=a \left(1+m/2\bar{r}\right)^2\bar{r}$
is the areal radius.
From \eqref{eq:def_ms_mass},
one can show that the Misner-Sharp mass for the McVittie spacetime is
(see e.g. \cite{faraoni2015cosmological})
\begin{align}
   \label{eq:msmass_mcvittie}
   M_{MS}
   =
   m_H
   +
   \frac{1}{2}H^2r_A^3
   .
\end{align}
Here $H\equiv\frac{1}{a(t)}\frac{da(t)}{dt}$ is the asymptotic Hubble expansion. 
When $H$ is constant, this is the Schwarzschild-de Sitter metric in coordinates 
analogous to outgoing Eddington-Finkelstein coordinates.
This metric is a solution to the Einstein equations provided $H$ satisfies the
Friedmann equation
\begin{align}
   H(t)^2
   =
   \frac{8\pi}{3}\rho(t)
   ,
\end{align}
where $\rho(t)\equiv \hat{\tau}^{\alpha} \hat{\tau}^{\beta}T_{\alpha\beta}$ is the
energy density of background fluid and $\hat{\tau}_{\alpha}$ the unit timelike normal 
to hypersurfaces of constant $t$. 
In principle, one may consider arbitrary FLRW backgrounds generated by cosmic
fluids satisfying any equation of state. 
However, the McVittie spacetime only describes a \emph{black hole}
embedded in a cosmology if the spacetime asymptotes to a de Sitter
background as $r\to\infty,t\to\infty$ \cite{Kaloper:2010ec}.
With this caveat in mind, from eq.~\eqref{eq:msmass_mcvittie}
we can think of $m_H$ as the mass of the black hole, and
$\frac{1}{2}H^2r_A^3=(4\pi r_A^3/3)\rho$ as the mass of the cosmological
fluid encapsulated within a sphere of radius $r_A$.
%=============================================================================
\subsection{Apparent horizons and event horizons
   in the McVittie spacetime\label{sec:dynamical_horizons_mcvittie}
}
   Here we briefly review different notions of horizons
in the McVittie spacetime \cite{Kaloper:2010ec,Faraoni:2012gz}
for reference and comparison to our numerical study.
To do so, we rewrite the McVittie line element \eqref{eq:mcvittie_metric0} in terms of
the areal radius such that the line element becomes
\begin{align}
   \label{eq:mcvittie_metric}
   g_{\alpha\beta}dx^{\alpha}dx^{\beta}
   =&
   -
   \left(1-\frac{2m_H}{r_A}-H^2r_A^2\right)dt^2
   -
   \frac{2Hr_A}{\sqrt{1-\frac{2m_H}{r_A}}}dtdr_A
   +
   \frac{1}{1-\frac{2m_H}{r_A}}dr_A^2
   +
   r_A^2d\Omega^2
   ,
\end{align}

We now use the notions introduced in section~\ref{sec:DH} to derive the location 
of the black hole and cosmological apparent horizons in McVittie spacetimes 
given by \eqref{eq:mcvittie_metric}.
We define the following orthonormal timelike and spacelike vectors, 
(that is $\hat{r}_{\alpha}\hat{r}^{\alpha}=1, 
\hat{\tau}^{\alpha}\hat{r}_{\alpha}=0$):
\begin{align}
   \hat{\tau}_{\alpha}dx^{\alpha}
   \equiv&
   \left(1-\frac{2m_H}{r_A}\right)^{1/2}dt
   ,\\
   \hat{r}_{\alpha}dx^{\alpha}
   \equiv&
   -
   Hr_A dt
   +
   \left(1-\frac{2m_H}{r_A}\right)^{-1/2}dr_A
   .
\end{align}
With the unit timelike vector $\hat{\tau}_{\alpha}$ and unit spacelike vector 
$\hat{r}_{\alpha}$ we can define the following metrics
\begin{align}
   h_{\alpha\beta}
   \equiv
   g_{\alpha\beta}+\hat{\tau}_{\alpha}\hat{\tau}_{\beta}
   ,\qquad
   \tilde{q}_{\alpha\beta}
   \equiv
   g_{\alpha\beta}+\hat{\tau}_{\alpha}\hat{\tau}_{\beta}-\hat{r}_{\alpha}\hat{r}_{\beta}
   ,
\end{align}
The tensor $h_{\alpha\beta}$ can be identified with the spatial
Riemannian metric of constant $t$ slices,
and $\tilde{q}_{\alpha\beta}$ can be identified with the angular metric.
In the coordinates eq.~\eqref{eq:mcvittie_metric}, 
the null expansions \eqref{eq:expansion} 
associated with the null vectors \eqref{eq:null_vectors} reduce to 
\begin{align}
   \Theta_{(l)}
   =
   \frac{1}{r_A} l^{\alpha} \partial_{\alpha} r_A
   =
   \frac{1}{r_A} l^{\hat{r}}
   =
   \frac{1}{r_A}\left(
      Hr_A + \sqrt{1-\frac{2m_H}{r_A}} 
   \right), \\
   \Theta_{(n)}
   =
   \frac{1}{r_A} n^{\alpha} \partial_{\alpha} r_A
   =
   \frac{1}{r_A} n^{\hat{r}}
   =
   \frac{1}{r_A}\left(
      Hr_A - \sqrt{1-\frac{2m_H}{r_A}} 
   \right)
   .
\end{align}
At the apparent horizons we have $\Theta_{(l)}\Theta_{(n)} = 0$.
For the McVittie solution the apparent horizons are then located
at the zeros of
\begin{align}
\label{eq:condition_thetapm_eq_0}
   H(t)^2r_A^3
   - 
   r_A
   + 
   2m_H
   =
   0
   .
\end{align}
In general, there are at most two real solutions.
The smaller root $r_{-} = r\left(\Theta_{(l)}=0,\Theta_{(n)}<0\right)$ 
is called the black hole apparent horizon, 
since it reduces to the Schwarzschild horizon
$r_{-} = 2 m_H$ in the limit where there is no background expansion $H\to 0$, 
while the larger root is called the cosmological apparent horizon,
as it reduces to the static de Sitter horizon 
$r_{+} = r\left(\Theta_{(n)}=0,\Theta_{(l)}>0\right) = 1/H_0$ in the limit 
$m \to 0$ and $H = H_0 > 0$ 
\cite{Kaloper:2010ec,Faraoni:2012gz,faraoni2015cosmological}.
In \cite{Kaloper:2010ec}, the authors further showed that the surface
defined by $\Theta_{(l)}=0$, $t\to0$ in fact defines the surface of
the event horizon for a black hole, provided that $\lim_{t\to\infty}H=H_0>0$
and $H>0$ for all $t$.
In this limit the black hole asymptotes to a Schwarzschild de-Sitter solution
as $t\to\infty$.
However, when the spacetime asymptotes to an FLRW cosmology with 
a scale factor obeying a power
law, so that $\lim_{t\to\infty}H=0$, then the authors
showed that the surface
defined by $\Theta_{(l)}=0$ asymptotes to a region with a weak
curvature singularity, essentially
due to the divergence of the radial pressure (together with the Ricci scalar)
required to keep the matter density on $t=const.$ slice constant .\\

Although the Schwarzschild-de Sitter black hole does not change size,
it is not unreasonable to expect that in more general FLRW spacetimes, 
black holes could expand or contract in size.
In particular, if one were to relax the no-accretion condition, 
then this would allow for the black hole to accrete matter 
from the surrounding cosmic fluid.
There are many generalizations of the McVittie spacetime 
in the literature and we refer the 
interested reader to \cite{faraoni2015cosmological} for a review.
However most generalizations of this spacetime are limited
to either a no-flux condition, 
or to specific kinds of matter fields.
For example, to work around
the no-flux condition, in \cite{Faraoni:2007es} the authors
had to use a fluid model that includes a ``heat'' current vector.
Because of this, it may be difficult to draw general conclusions 
from the generalized McVittie models.
For example, there are conflicting claims about whether 
a black hole expands/contracts  
in a universe filled with matter/phantom energy, where studies making
use of the McVittie model reach the opposite conclusion of studies that 
do not make use of the model
\cite{Babichev:2004yx,Gao:2008jv,Gao:2011tq}.

%=============================================================================
\bibliography{main}
\bibliographystyle{JHEP}

\end{document}